\documentclass[twocolumn,longbib]{aastex7}
\usepackage{xurl}
\usepackage{makecell}
\usepackage{booktabs}
\usepackage{amsmath}
\usepackage{float}
\received{December 8, 2025}
\revised{January 29, 2026}
\accepted{February 12, 2026}
\submitjournal{PSJ}

\shorttitle{}
\shortauthors{Proudkii et al.}

\begin{document}

\title{Vertical Temperature Structure in Io's Atmosphere from ALMA SO$_2$ Observations}

\correspondingauthor{Timothy N. Proudkii}
\email{timproudkii@caltech.edu}

\author[orcid=0009-0005-1065-4109]{Timothy N. Proudkii}
\affiliation{Division of Physics, Mathematics, and Astronomy, California Institute of Technology, Pasadena, CA 91125, USA}
\email{timproudkii@caltech.edu}

\author[orcid=0000-0002-9068-3428]{Katherine de Kleer}
\affiliation{Division of Geological and Planetary Sciences, California Institute of Technology, Pasadena, CA 91125, USA}
\email{dekleer@caltech.edu}

\author[orcid=0000-0002-4278-3168]{Imke de Pater}
\affiliation{Department of Astronomy, Department of Earth and Planetary Science, University of California, Berkeley, CA 94720, USA}
\email{imke@berkeley.edu}

\author[orcid=0000-0002-8178-1042]{Alexander E. Thelen}
\affiliation{Division of Geological and Planetary Sciences, California Institute of Technology, Pasadena, CA 91125, USA}
\email{athelen@caltech.edu}

\author[orcid=0000-0001-9867-9119]{Statia Luszcz-Cook}
\affiliation{New York University, New York, NY 10003, USA}
\affiliation{American Museum of Natural History, New York, NY 10024, USA}
\email{shc7964@nyu.edu}

\author[orcid=0000-0001-7168-1577]{Emmanuel Lellouch}
\affiliation{LIRA, Observatoire de Paris, Université PSL, CNRS, Sorbonne Université, Université Paris Cité, 5 place Jules Janssen, 92195 Meudon, France}
\email{Emmanuel.Lellouch@obspm.fr}

\author[orcid=0000-0002-9820-1032]{Arielle Moullet}
\affiliation{National Radio Astronomy Observatory, Charlottesville, Virginia, USA}
\email{amoullet@nrao.edu}

\begin{abstract}
The structure of Io's atmosphere is controlled by competing processes, from volcanic outgassing and sublimation to radiative cooling and plasma heating. Yet, the lack of an observationally-derived temperature profile has left this balance unconstrained. We used four epochs of Atacama Large Millimeter/submillimeter Array (ALMA) Band 7 (275-373 GHz) and Band 8 (385-500 GHz) SO$_2$ spectroscopy to retrieve Io's vertical atmospheric temperature profiles. To mitigate longstanding degeneracies common in atmospheric retrievals, we performed a simultaneous multi-line analysis combined with line-of-sight disk-resolved Doppler velocity maps and a forward model that included a sub-beam velocity-dispersion term. This modeling approach enabled the separation of thermal and dynamical line-shape contributions. On the leading hemisphere, we retrieved a cold, quasi-isothermal lower atmosphere ($\sim$124-137~K up to $\sim$0.5~nbar), followed by a thermospheric rise reaching hundreds of kelvins by $\sim10^{-2}$~nbar. On the trailing hemisphere, our fits yielded qualitatively similar profiles but consistently retrieved lower SO$_2$ column densities. The lower column \added{densities} confined line formation to the first few kilometers, making the trailing hemisphere spectra statistically consistent with an isothermal atmosphere. Across datasets, we retrieved fractional \added{gas} coverages of $\sim$35-50$\%$ and sub-beam velocity dispersions of $\sim$25-85$\mathrm{~m~s^{-1}}$, \added{encoding line-of-sight velocity dispersion within a beam element in excess of the disk-resolved Doppler velocity map.} Together, these retrievals deliver the first vertically resolved temperature profiles of Io's atmosphere, reveal robust vertical structure on the dayside leading hemisphere, and offer new constraints on Io's thermal energy balance.
\end{abstract}



\section{Introduction} 
For the past $\sim$4.5 billion years, a hellish landscape of fire and ice has defined Io---Jupiter's innermost Galilean moon---where volcanic eruptions and sulfurous plumes shape a dynamic atmosphere. Its surface exhibits vivid reds and yellows largely produced by plume deposits of sulfur, which may reflect different allotropes and temperature-dependent phases, while the patchy white spots that blanket the surface are produced by sulfur dioxide (SO$_2$) frost deposits \citep{Sagan1979, Johnson1979, Bertaux1979}. This kaleidoscope of a terrain is constantly reshaped by widespread volcanic activity, which is driven by strong tidal forces from Jupiter and its neighboring moons, Europa and Ganymede. These gravitational interactions have sustained continuous volcanic activity on Io since its formation \citep{deKleer2024}.

Beyond reshaping Io's surface, this frequent volcanic activity also plays a crucial role in sustaining its atmosphere. Io's atmosphere is in perpetual turnover, as it is continuously lost to space primarily by ion-neutral collisions between energetic particles from Jupiter's intense magnetosphere and molecules in Io's atmosphere \citep{Bengal2020}. Estimates suggest that Io's mass loss rate is 1-3 ton s$^{-1}$ \citep{Spencer1996, dols2008multispecies}. Its atmosphere is then replenished through two primary mechanisms: volcanic outgassing and sublimation of surface SO$_2$ frost. Volcanic plumes inject fresh gas directly into the atmosphere \citep{Spencer1996}, while solar heating drives the sublimation of SO$_2$ frost deposits. Sublimation is generally considered the dominant source of Io's atmosphere \citep{Tsang2012, tsang2013sublimation, Bengal2020, dePater2023, dott2025observed, Giles2024}, but its contribution varies with latitude, time of day, and Io's position in eclipse, when rapid cooling causes the atmosphere to partially collapse \citep{tsang2016collapse, dePater2020b}. However, Io's volcanism is a fundamental influence on both Io's atmosphere and surface, as all atmospheric SO$_2$ ultimately derives from volcanic activity, which supplies both the plumes and the frost reservoir \citep{Lellouch2005review}. Additionally, at certain times and locations, volcanic outgassing can certainly dominate the local atmospheric supply \citep{dePater2020b}. The balance between these replenishment and loss processes dictates the structure and variability of Io's atmosphere, making it highly dynamic and dependent on both internal and external mechanisms.

\subsection{Hemispheric Asymmetries and Plasma Interactions}\label{subsec: theory-models}
As Io orbits Jupiter, the atmosphere undergoes dramatic changes, alternating between a ``daytime'' state, where SO$_2$ frost is sublimated by solar heating, and a ``nighttime'' state, where SO$_2$ gas condenses onto the surface back into frost. \added{This condensation happens on time scales possibly as short as $\sim$70 seconds \citep{summers1996photochemistry, dePater2002} and observations have shown the atmosphere to collapse within a few minutes \citep{tsang2016collapse, dePater2020b}}. However, as non-condensable gases accumulate near the surface they can form a diffusive layer that may slow any additional collapse, and this layer can take tens of minutes to develop \citep{Moore2009}. The atmosphere, when reemerging in sunlight, is then reformed after $\sim\!10$~minutes \citep{dePater2020b}. This behavior illustrates the extreme variability of Io's atmosphere. 

\added{Io's tidally locked nature further contributes to its atmospheric extremes, with a permanent sub-Jovian and anti-Jovian hemisphere. Observations have shown that the lower atmosphere exhibits pronounced asymmetries: the sub-Jovian side generally has significantly lower SO$_2$ column densities than the anti-Jovian side \citep{McGrath2000, mcgrath2004satellite, jessup2004atmospheric, Spencer2005, lellouch2007io, moullet2008first, Tsang2012, Giles2024}. This asymmetry, in large part, can be explained by differences in solar heating, since the sub-Jovian side is more frequently in eclipse and therefore exhibits a lower surface temperature than the anti-Jovian side (reducing SO$_2$ sublimation). Although the lower level of volcanic activity on the sub-Jovian side may contribute, modeling work shows that the observed hemispherical asymmetry is primarily driven by the eclipse effect \citep{walker2012parametric}. At the same time, Io's synchronous rotation also defines a permanent ``leading'' hemisphere facing forward in its orbit and a ``trailing'' hemisphere facing backward. As a result, the trailing hemisphere is persistently exposed to a dense, co-rotating plasma torus that sweeps past Io at a relative speed of $\sim$$57~\mathrm{km~s^{-1}}$ \citep{Wong2000}. This preferential hemispheric plasma bombardment has dramatic consequences: notably, models predict that it serves as both a mechanism for gas escape \citep{Smyth1998} and as a dominant heating source (especially in the upper atmosphere; \citealt{Johnson1989, Strobel1994}).}

\begin{figure*}
    \centering
    \includegraphics[width=\textwidth]{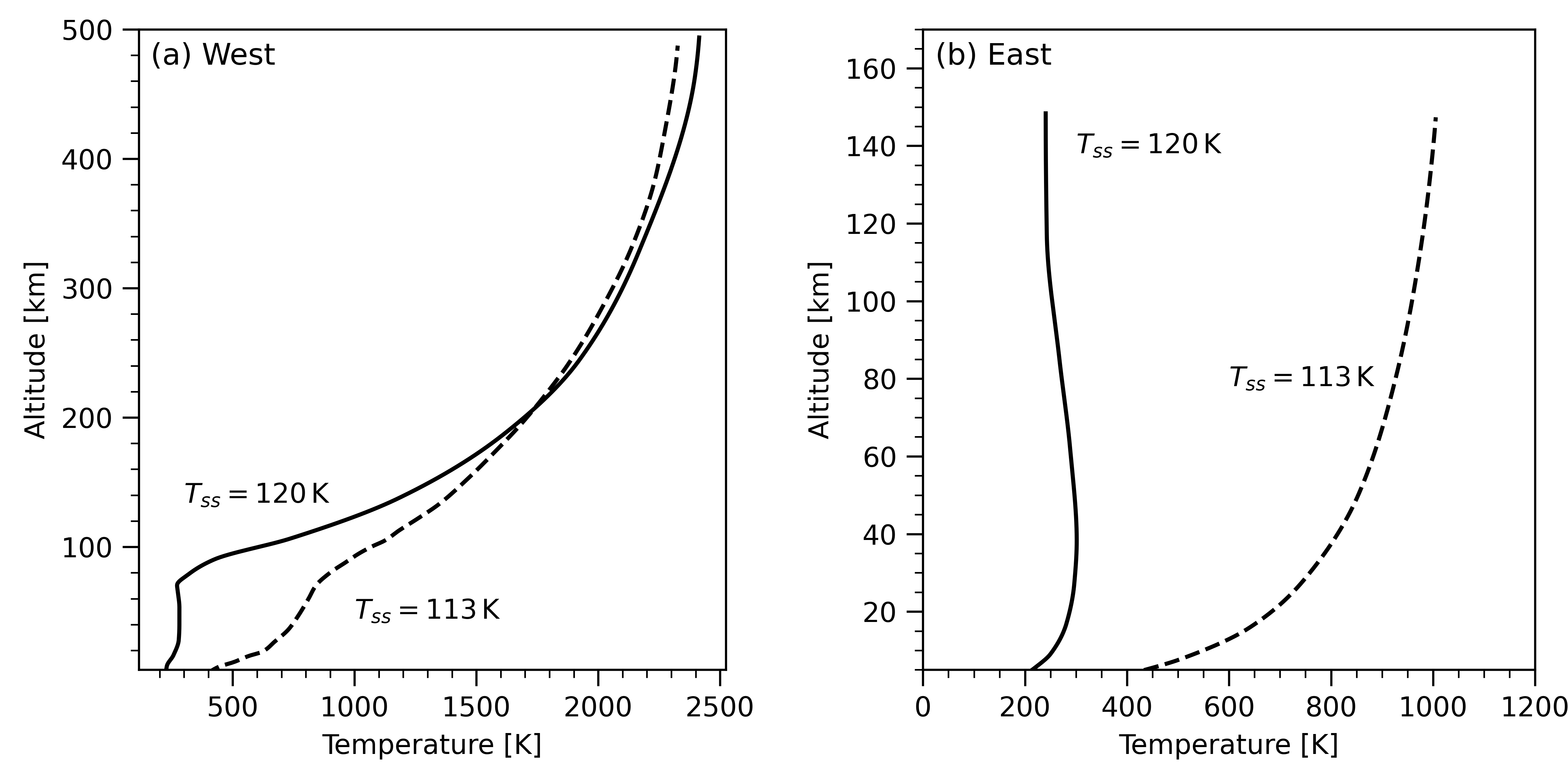}
    \caption{Reproduction of Figure 3 from \citet{Wong2000}, showing modeled vertical temperature profiles at the subsolar point for high-density ($T_\mathrm{ss} = 120$~K) and low-density ($T_\mathrm{ss} = 113$~K) atmospheres. Here, \( T_\mathrm{ss} \) denotes the subsolar surface temperature, which sets the boundary condition for SO$_2$ sublimation. Panel (a) corresponds to western elongation (Io's sunlit trailing hemisphere), where plasma heating on the dayside leads to a more extended atmosphere. Panel (b) shows eastern elongation (Io's sunlit leading hemisphere), where reduced dayside plasma heating results in a cooler, more compact atmosphere.}
    \label{fig:wong_model}
\end{figure*}

To disentangle the effects of this preferential plasma bombardment on Io's trailing hemisphere, it is critical to obtain direct measurements of the vertical temperature structure of Io's atmosphere. As shown in \citet{Wong2000}, plasma collisional heating drives variations in exobase altitude and temperature. \citet{Wong2000} performed numerical calculations for Io's sublimation-driven atmosphere at both eastern and western elongations---corresponding to Io's sunlit leading and trailing hemispheres, respectively---using an improved version of the multi-species hydrodynamic model originally developed by \citet{WongJohn1996}. Their model predicts that for Io's sunlit trailing hemisphere, the atmosphere is more extended and exobase temperatures can exceed 2,400~K. Additionally, depending on the SO$_2$ density, \citet{Wong2000} showed that the atmosphere remains quasi-isothermal up to about 70~km, increasing in temperature sharply afterwards. In contrast, for Io's sunlit leading hemisphere, their model predicts a less extended atmosphere with temperatures of 200~K at the exobase. Additionally, their modeled temperature profiles remain quasi-isothermal at high SO$_2$ densities, and monotonically increasing at lower densities. This substantial difference in thermal structure across sunlit hemispheres underscores the asymmetry introduced by plasma heating and the importance of resolving vertical atmospheric properties. Figure~\ref{fig:wong_model} shows the modeled subsolar temperature profiles that illustrate these predicted differences in atmospheric extent and exobase temperature. Notably, their model predicts a quasi-isothermal lower layer followed by a thermospheric warming for Io's sunlit trailing hemisphere when the atmosphere is dense enough, as shown in Figure~\ref{fig:wong_model}~(a).

The predicted vertical structure from \citet{Wong2000} builds on earlier models that had already suggested such a feature in Io's lower atmosphere. \citet{lellouch1990io}, using an aeronomic model developed for Titan but adopted for Io, showed that if SO$_2$ surface pressures were high, non-LTE cooling in the $\nu_1$, $\nu_2$, and $\nu_3$ vibrational bands of SO$_2$ could generate a temperature minimum (a mesopause) within the first scale height. Extending this work, \citet{Lellouch1992} emphasized that SO$_2$ non-LTE vibrational cooling scales strongly with density ($\propto P^2$), and hypothesized that this cooling could modify the lower atmosphere thermal structure by generating a quasi-isothermal structure from the surface up to the nanobar level (depending on the SO$_2$ density). \citet{Strobel1994} then developed a full radiative-conductive model incorporating solar, Joule, and plasma heating as well as non-LTE rotational and vibrational cooling. \citet{Strobel1994} demonstrated that for surface pressures greater than $\sim\!10$~nbar, Io's atmosphere should host a quasi-isothermal atmosphere up to a pressure of $\sim\!1$~nbar. Then, at lower atmospheric pressures (higher altitudes), Joule heating takes over and drives a rapid temperature increase, reaching $\sim\!1,000$~K by the 0.1-nbar level at $\sim\!200$~km (depending on the assumed Pedersen conductance of Io's ionosphere). Interpretation of recent James Webb Space Telescope (JWST) sulfur lines (1.08 $\&$ 1.13 micrometers) from \citet{dePater2025first} require atmospheric temperatures of order 1,700 K above $\sim\!$200-300 km, in agreement with \citet{Strobel1994}'s upper atmospheric temperatures.

A well-constrained vertical temperature profile is therefore essential not only for testing the validity of these model predictions but also for decoupling the roles of solar heating, SO$_2$ cooling, Joule heating, and plasma collisional heating. Each of these heating mechanisms is expected to dominate at different altitudes: solar heating and SO$_2$ cooling primarily affects the lower atmosphere near the surface, Joule heating can occur throughout the atmosphere depending on conductivity, and plasma collisional heating tends to deposit energy in the upper atmosphere. A vertical temperature profile would thus serve as a key diagnostic for interpreting Io's atmospheric structure and for constraining the balance between atmospheric heating and cooling processes.

\subsection{Limitations in Retrieving Io's Atmospheric Temperature}\label{subsec: Temp-limitations}
Despite extensive observations across multiple wavelengths, key aspects of Io's atmospheric structure---particularly its vertical temperature profile---remain poorly constrained. While various observational techniques have provided valuable insights into Io's atmospheric properties, each method has inherent limitations that hinder the ability to fully characterize its thermal structure. 

At \textit{Ultraviolet/Visible (UV-visible)} wavelengths, constraints on Io's atmospheric temperature come from UV spectroscopy of SO$_2$ absorption bands, but the diagnostic is intrinsically weak. The temperature sensitivity arises from subtle skewness in band shapes, and published UV analyses have historically inferred or assumed a wide range of 110-500~K \citep{ballester1994detection, McGrath2000, spencer2000discovery, jessup2004atmospheric, jessup2007sulfur, tsang2013synergistic, jessup2015spatially}. As emphasized by \citet{Lellouch2015}, the most recent and highest-quality observations narrow the allowed regime to colder values, with \citet{jessup2004atmospheric} favoring 150-250~K and \citet{tsang2013synergistic} and \citet{jessup2015spatially} indicating 100-200~K. In all cases, the UV-derived temperatures are line-of-sight (LOS) averages over the SO$_2$ distribution rather than vertically resolved profiles and remain sensitive to assumptions about SO$_2$ cross-sections, band transmission, and column density.

Observations of SO$_2$ ro-vibrational lines in the thermal \textit{mid-infrared} have been made but are strongly affected by non-LTE (local thermodynamic equilibrium) effects, as radiative de-excitation dominates over collisional processes \citep{Lellouch1992, Spencer2005}. In LTE, molecular energy levels are populated according to the Boltzmann distribution, meaning that line intensities provide a direct measure of kinetic temperature. However, in non-LTE conditions, the vibrational and kinetic temperatures decouple, leading to line intensities that do not reliably reflect Io's true atmospheric thermal structure. In the \textit{mid-infrared}, only absorption lines have been detected, indicating vibrational temperatures colder than Io's surface temperature. While kinetic temperature could---in principle---be inferred below the altitude where \citet{Gratiy2010} establish rotational LTE (about 50~km in their adopted profile, corresponding to pressure $P\sim0.02$~nbar), this process is complicated by optical depth effects, variations in SO$_2$ column density, and radiative transfer uncertainties \citep{Tsang2012, dePater2023}. Observations from the 19~$\mu$m band indicates temperatures below 150~K \citep{Spencer2005}, with the anti-Jovian hemisphere reaching as low as 108 $\pm$ 18~K \citep{Tsang2012}.

In the \textit{near-infrared}, SO$_2$ gas has been detected at 4~$\mu$m, where the lines are formed mostly in the solar reflected component. The isotherm temperature estimate of 170~$\pm$~20~K derived by \citet{Lellouch2015} comes from the relative distribution of line depths across multiple rotational transitions in the 4~$\mu$m region. This analysis effectively constitutes a rotational-temperature determination, which remains valid even when the individual lines are not spectrally resolved. By contrast, \textit{mid-infrared} 19~$\mu$m observations indicate temperatures below 150~K. The discrepancy between these two wavelength regimes underscores the difficulty of constraining Io's atmospheric temperatures, as different wavelength regimes probe different excitation conditions and different regions in the atmosphere.

These challenges across wavelengths highlight the need for an alternative approach that directly measures kinetic temperature in LTE conditions. \textit{Millimeter/Submillimeter (mm/sub-mm)} observations provide this alternative by probing high-pressure SO$_2$ rotational transitions, which remain in LTE \added{at pressures greater than $P\sim0.02$~nbar (reported by \citet{Gratiy2010} as about 50~km in their adopted profile). Because the pressure-altitude mapping is model dependent, we treat the pressure threshold as the more transferable statement of the LTE regime. Because LTE couples the rotational level populations to the local kinematic temperature, this regime} allows for direct retrieval of atmospheric temperatures by analyzing the widths and intensities of SO$_2$ emission lines. However, at higher altitudes, the assumption of LTE no longer holds---making \textit{mm/sub-mm} techniques less effective for probing the upper atmosphere. 

Even for the lower atmosphere that remains in LTE, temperature inferences from \textit{mm/sub-mm} spectroscopy have historically been difficult to reconcile. Early analyses that assumed the SO$_2$ line widths were purely thermal produced unrealistically high temperatures, in some cases approaching 600~K \citep{Lellouch1992}. These interpretations are now understood to be incorrect because they did not account for velocity broadening from winds and plume dynamics. Once atmospheric dynamics were incorporated, both plume models \citep[e.g.][]{lellouch1996urey} and subsequent wind-corrected analyses of mm data \citep{moullet2008first, Moullet2010, moullet2013exploring} converged to much colder values, typically 130-300~K. More recent analyses from \citet{Roth2020, dePater2020b, deKleer2024} agree with this temperature range. The difficulty in determining Io's atmospheric temperature in the \textit{mm/sub-mm} arises because SO$_2$ line shapes reflect a superposition of thermal Doppler broadening, sublimation-driven winds, and localized volcanic plumes, all of which imprint Doppler shifts and asymmetries that complicate temperature retrievals. Another limitation is that previous studies were not able to disentangle the vertical dependence of these effects and, therefore, only reported a single characteristic temperature---representing a weighted average over multiple altitudes rather than a true vertical temperature profile. Without multi-line observations capable of differentiating temperature variations with altitude, these measurements fail to resolve the underlying thermal structure of Io's atmosphere.

Achieving reliable, vertically resolved temperature profiles on Io therefore demands a combination of observational and modeling advances. First, multi-transition observations spanning lines with a range of lower-state energies are needed so that different altitudes in the lower atmosphere are sampled. Second, high spectral resolution is required to separate intrinsic thermal broadening from instrumental and dynamical contributions. Third, high spatial resolution is required to reduce the smearing of Io's highly variable atmospheric state across the disk. Finally, contemporaneous LOS Doppler velocity maps---derived from the same data---are needed to prescribe large-scale Doppler shifts and isolate any remaining line broadening as truly sub-beam velocity dispersion. By fitting all transitions simultaneously in a Bayesian framework that includes both thermal Doppler and a free wind-dispersion term, we can significantly reduce and better understand the degeneracies between temperature, column density, and non-thermal motions---enabling a more robust inference of Io's vertical temperature structure. 

\subsection{Outline}
In this study, we used high spatial and spectral resolution Atacama Large Millimeter/submillimeter Array (ALMA) observations of SO$_2$ rotational emission lines to retrieve vertically resolved temperature profiles in Io's atmosphere, both for its sunlit leading and trailing hemispheres. We incorporated LOS Doppler velocity maps and dispersion into the forward model to decouple thermal and wind broadening in the observed line shapes, tested different temperature structures of varying complexity, employed Markov chain Monte Carlo (MCMC) sampling to map the joint posterior of the atmosphere model parameters, and evaluated model complexity using leave-one-out cross-validation. 

Following this section, Section~\ref{sec:Observations and Data Reduction} describes the ALMA observational parameters and data reduction procedures. Section~\ref{sec:Modeling Framework} outlines the radiative transfer model, LOS Doppler velocity map implementation, temperature parameterizations, and the MCMC retrieval method. Section~\ref{sec:Results} presents the retrieved vertical temperature profiles for both the leading and trailing hemispheres. In Section~\ref{sec:Discussion}, we compare these results to theoretical predictions and assess the impact of different heating/cooling mechanisms. Section~\ref{sec:Conclusion} summarizes the key findings and highlights implications for future observations of Io's atmosphere.

\section{Observations and Data Reduction}
\label{sec:Observations and Data Reduction}
\begin{deluxetable*}{ccccccccc}  
\tablecaption{Observational Parameters \label{tab:obsparams}}
\tablehead{
\colhead{Project} & 
\colhead{Dataset} &
\colhead{Date} & 
\colhead{Ang. Dia.\tablenotemark{a}} & 
\colhead{Ang. Res.} & 
\colhead{Spec. Res.} & 
\colhead{Lat.} & 
\colhead{W. Lon.} &
\colhead{Targeted Hemi.}\\ [-5pt] 
\colhead{Code \#} & 
\colhead{} & 
\colhead{} & 
\colhead{[\arcsec]} & 
\colhead{[\arcsec]} & 
\colhead{[kHz] [km s$^{-1}$]} & 
\colhead{[deg]} & 
\colhead{[deg]} &
\colhead{}
}
\startdata
2019.1.00216.S & J21-L & 2021 Jul 15 & 1.20 & 0.20 $\times$ 0.20 & 141 [0.12] & 0.86 &  98.0 & Leading  \\
               & J21-T & 2021 Jul 23 & 1.22 & 0.20 $\times$ 0.20 & 141 [0.12] & 0.86 & 266.0 & Trailing \\
2021.1.00849.S & M22-T & 2022 May 18 & 0.92 & 0.30 $\times$ 0.27 & 244 [0.17] & 2.09 & 277.1 & Trailing \\
               & M22-L & 2022 May 24 & 0.94 & 0.35 $\times$ 0.23 & 244 [0.17] & 2.14 &  76.8 & Leading  \\
\enddata
\tablenotetext{a}{This column lists the angular diameter of Io at the time of each observation. Data from ALMA Project Code \#2019.1.00216.S and \#2021.1.00849.S were used in \citet{Thelen2025}, while only data from \#2021.1.00849.S were used in \citet{deKleer2024}.}
\end{deluxetable*}

We analyzed two ALMA data sets each obtained in ALMA's Band 7 (275-373 GHz; 0.8-1.1 mm) in July 2021 and Band 8 (385-500 GHz; 0.6-0.8 mm) in May 2022, comprising a total of four individual observations of Io: two targeting the sunlit leading hemisphere and two the sunlit trailing hemisphere. These observations were previously presented by \citet{Thelen2025} (Band 7) and by \citet{deKleer2024} (Band 8). Observational parameters for all four epochs are summarized in Table~\ref{tab:obsparams}. The resulting image cubes were adopted from these previous publications without further modification. 

The original calibration and imaging of these ALMA observations were performed using the Common Astronomy Software Applications (CASA) package \citep{McMullin2007, Casa2022}, following standard ALMA Observatory pipeline scripts. The continuum was removed in the $uv$-plane using CASA's \texttt{uvcontsub} and subsequent imaging was performed using CASA's \texttt{tclean} deconvolution algorithm \citep{Hogbom}. Imaging was carried out using Briggs weighting with a robust parameter of 0.5, resulting in the synthesized beam sizes tabulated in Table~\ref{tab:obsparams}, which depend on the array configuration. Pixel scales of $0.''01$ (Band 7) and $0.''03$ (Band 8) were used to adequately sample the beam. Imaging included iterative phase self-calibration, following procedures described in \citet{dePater2020b} and \citet{deKleer2024}. Full calibration and imaging details are given in \citet{dePater2020b, Thelen2025, deKleer2024}. The native spectral resolution is 141~kHz (0.12~km~s$^{-1}$) for the Band 7 data and \added{244~kHz (0.17~km~s$^{-1}$)} for the Band 8 data. Throughout this manuscript, we refer to the July 2021 Band 7 data as J21-L/T and the May 2022 Band 8 data as M22-L/T, with `L' and `T' indicating leading or trailing hemisphere, respectively.

From these calibrated and cleaned image cubes, we extracted disk-center spectra with an aperture corresponding to roughly one ALMA beam. This choice maximized signal-to-noise in a single resolution element while keeping our LOS as close to perpendicular to the surface as possible. This choice is favorable since the path through the atmosphere is mostly vertical at disk-center; therefore, horizontal gradients in brightness or wind-induced Doppler shifts are minimized. Additionally, since Io's atmosphere is highly variable and spatially heterogeneous, our choice of a disk-center aperture minimizes mixing of signals from regions that may be different in pressure, temperature, or wind speed. By isolating a single resolution element at each epoch, our spectra capture near-instantaneous snapshots of the local atmosphere with minimal spatial or diurnal averaging that would wash out rapid changes driven by Io's extreme environment---effects that become more difficult to interpret when extracting spectra over Io's full disk.

We tested the robustness of our choice in spectral extraction region by repeating the retrieval on spectra extracted with two additional apertures: a circular aperture approximately twice the diameter of the ALMA beam, and an aperture encompassing Io's full disk. The retrieved temperature profiles showed minimal quantitative differences ($\lesssim$10-20~K at pressure levels our data are sensitive to), and preserved the same overall shape. Given the consistency of the retrieved profiles across our tested apertures, the reduced complexity of the geometry, and the lower computational cost, we adopt the ALMA beam-sized, disk-center extraction for all subsequent analysis.

We calculated the flux-density uncertainty, $\sigma$, for each spectral window by taking the standard deviation of the disk-center spectrum in channels that showed no emission (using the same beam-sized aperture). This $\sigma$ captured Io's continuum thermal noise and was then adopted as the uncertainty for every data point in that line. We verified that the noise level does not vary spatially outside Io's disk in the cleaned image cubes, indicating that the interferometric imaging does not introduce position-dependent noise structure.

We targeted a suite of SO$_2$ rotational transitions that sample a range of line strengths and excitation conditions. In Table~\ref{tab:so2_lines} we tabulate the rest frequency, line strength at 300 K, and lower-state energy ($E_{\rm low}$) for each line.

\begin{deluxetable*}{ccc}
\tablecaption{Targeted SO$_2$ Rest Frequencies\label{tab:so2_lines}}
\tablehead{
  \colhead{Frequency} & 
  \colhead{Line strength, $S$} & 
  \colhead{$E_{\mathrm{low}}$} \\[-4pt]
  \colhead{[GHz]} & 
  \colhead{[$\mathrm{cm^{-1}/mol/cm^{2}}$]} & 
  \colhead{[cm$^{-1}$]}
}
\startdata
\tableline 
\multicolumn{3}{c}{\text{Dataset J21-L \& J21-T}} \\
\tableline
358.215640 & 1.10199e$^{-21}$ & 116.862 \\
346.652167 & 1.02655e$^{-21}$ & 105.299 \\
346.523878 & 5.71320e$^{-22}$ & 102.750 \\
359.151156 & 4.63638e$^{-22}$ & 211.076 \\
360.290400 & 3.66167e$^{-22}$ & 413.312 \\
348.387796 & 3.09371e$^{-22}$ & 191.843 \\
349.783318 &  1.14309e$^{-22}$ & 732.935 \\
\tableline 
\multicolumn{3}{c}{\text{Dataset M22-T}} \\
\tableline
430.2286454 & 1.53454e$^{-21}$ & 168.493 \\
430.2323107 & 9.47282e$^{-22}$ & 138.227 \\
430.1937035 & 9.75168e$^{-22}$ & 180.632 \\
416.8255538 & 6.35447e$^{-22}$ &  289.053 \\
419.0190334 & 5.15798e$^{-22}$ &  331.435 \\
418.8157956 & 9.58251e$^{-23}$ & 192.736 \\
429.8638418 & 1.77002e$^{-22}$ & 659.183 \\
\tableline 
\multicolumn{3}{c}{\text{Dataset M22-L}} \\
\tableline
430.1937035 & 9.75168e$^{-22}$ & 180.632 \\
416.8255538 & 6.35447e$^{-22}$ &  289.053 \\
419.0190334 & 5.15798e$^{-22}$ &  331.435 \\
429.8638418 & 1.77002e$^{-22}$ & 659.183 \\
418.8157956 & 9.58251e$^{-23}$ & 192.736 \\
\enddata
\tablecomments{Grouped by dataset (see Table~\ref{tab:obsparams}); J21-L and J21-T share the same line set. Line rest frequencies, strengths at 300 K, and $E_{\rm low}$ are adopted from the Cologne Database for Molecular Spectroscopy (CDMS; \citealt{CDMS2001, CDMS2016}).}
\end{deluxetable*}

\section{Spectral Line Modeling and Retrieval}
\label{sec:Modeling Framework}

We briefly summarize the retrieval workflow and how key parameter degeneracies were handled. First, we forward-modeled all the disk-center SO$_2$ line profiles with an LTE, hydrostatic radiative-transfer model that allowed flexible one-, two-, or three-node temperature profiles. \added{Throughout this work, ``nodes'' refer to discrete pressure levels at which the temperature is explicitly specified, with the full vertical profile obtained by linear interpolation between them.} The model also included a fractional beam-filling term for patchy emission, a sub-beam wind dispersion term for non-thermal line broadening, and ingested spatially resolved LOS Doppler velocity maps (Subsection~\ref{subsec: RT Model}). We then performed a single joint Bayesian fit across all lines to infer SO$_2$ column, node temperatures, fractional coverage, and sub-beam wind dispersion (Subsection~\ref{subsec: MCMC}). Parameter related degeneracies were mitigated primarily by the multi-line, multi-opacity fit and secondarily by physically motivated priors that gently \added{constrained} parameter ranges---hemisphere-dependent SO$_2$ column density ranges from long-term monitoring, a soft frost-equilibrium constraint, and simple bounds on temperatures, coverage, and wind dispersion (Subsection~\ref{subsec: Priors}). Finally, we compared our retrieved temperature profiles and report which is best supported by the data statistically (Subsection~\ref{subsec: PSIS}).

\subsection{Atmospheric Radiative Transfer Model}\label{subsec: RT Model}
To model the extracted, disk-center spectral line profiles, we built upon a radiative transfer (RT) code tailored to Io's lower atmosphere \citep{Code}. This model was first used and described in \citet{dePater2020b}, and subsequently (after translating to python) in \citet{deKleer2024}. We assumed hydrostatic equilibrium, which allowed us to derive atmospheric densities directly from an input vertical temperature profile (TP). Additionally, we assumed LTE throughout our modeled atmosphere, which is appropriate for the rotational transitions of SO$_2$ under Io's typical atmospheric conditions \citep{Lellouch1992, Gratiy2010}. 

As a robustness check, we tested the influence of non-LTE effects on our retrieval by constructing an atmospheric model that truncated the emitting layer at the level where the non-LTE source function becomes about 70$\%$ of the LTE source function. We used Figure~3 from \citet{Gratiy2010} to identify the corresponding density for this cutoff. From these tests, we found that this truncated retrieval yielded identical temperature profiles and other best-fit parameters because the contribution functions of the observed lines peak at pressures greater than $10^{-2}~\mathrm{nbar}$, so layers above the cutoff contribute negligibly to the emergent flux and do not strongly affect the fit. We therefore retain the LTE formulation in our RT model and limit our quantitative interpretation of the retrieved temperature profiles to pressures greater than $10^{-2}~\mathrm{nbar}$.

We modeled the TP using a flexible multi-node parameterization, which enabled us to test increasingly complex vertical temperature profiles. In this work, we explored models with one, two, and three temperature nodes to characterize Io's atmospheric TP. Our modeled atmosphere was discretized into 200 layers, evenly spaced in logarithmic pressure, spanning from the surface up to a pressure 10 scale heights above the surface. The surface pressure was dynamically calculated based on the total gas column density, which we computed from the SO$_2$ column plus small contributions from minor species and isotopologues. In the simplest one-node TP (1T), the atmosphere was isothermal, defined by a single temperature at all altitudes. In the two-node TP (2T), we allowed for a linear temperature gradient between a freely fit temperature at the first atmospheric layer and another at the top of the modeled atmosphere. The three-node TP (3T) introduced an intermediate node at a fitted transition pressure. This three-node TP allowed for more complex structures such as temperature inversions. Our simple, yet flexible, approach allowed us to fit the data without biasing the result by forcing functional forms of the TP (e.g., a low-order polynomial). 

In addition to these temperature profile parameterizations, we further adapted the RT code to ingest LOS Doppler velocity maps retrieved by \citet{Thelen2025}. These maps captured a mix of global and local wind phenomena, including broad day-to-night flows near the subsolar point and patchy equatorial inflows likely driven by spatially variable SO$_2$ frost sublimation. Io's solid-body rotation ($\sim\!75 \mathrm{~m~s^{-1}}$ at the equator) was also incorporated into the maps we use. The strongest winds aligned with Io's orbital motion and the direction of the co-rotating plasma torus. To avoid transition-specific artifacts, we adopted the average SO$_2$ LOS Doppler velocity map from 2-3 of the strongest lines in each observation. 

\added{These pixel-resolved velocity fields were applied directly to the RT calculation as fixed Doppler shifts prior to spatial integration. The observed spectra were already shifted to Io's rest frame prior to fitting, so no additional ALMA-Io LOS velocity is required.} Incorporating these spatially resolved velocity fields into the RT model allowed us to reproduce the expected Doppler shift in the line centroid arising from both rotation and large-scale winds. \added{Therefore, we did not include a separate free line-centroid parameter in the fit, reducing the number of free parameters and lowering covariance among the retrieved parameters.} The LOS Doppler velocity maps were treated as fixed inputs and their uncertainty was not propagated \added{because any residual uncertainty in the LOS velocity field does not behave like a global shift of the disk-center spectrum. Instead, map errors introduce small, spatially varying velocity mismatches across the aperture that primarily act as additional unresolved broadening once the pixels are summed. This effect is naturally absorbed by the fitted non-thermal dispersion parameter, $v_{\text{disp}}$, which we introduce in-depth shortly. Therefore, using the LOS Doppler velocity maps as fixed inputs does not meaningfully affect the disk-center line centroid or the retrieved temperature profile.}

The maps we use from \citet{Thelen2025} were derived directly from the same ALMA data sets analyzed in this study, which ensured consistency in observing geometry and atmospheric state. We computed model image cubes at several trial model pixel scales and selected the coarsest grid that did not significantly alter the disk-center integrated model spectrum. Based on this criterion, we adopted a model pixel scale of $0.''02$ for the J21-L/T dataset and $0.''09$ for the M22-L/T dataset. We then converted the modeled brightness to flux per pixel and summed over the same disk-center pixel mask used for the data to obtain the disk-center integrated model spectrum. We performed our RT calculations at a frequency resolution four times finer than that of the ALMA observations. These modeling choices reflect a balance between computational efficiency and RT accuracy. 

Because we already applied pixel-by-pixel LOS velocity shifts from the Doppler velocity maps, any leftover line broadening---beyond the thermal Doppler component and opacity effects---likely arose from velocity structure the maps don't capture (e.g., vertical shear across the line-forming layers, sub-beam variability, unresolved plume/outflow structure, or time variability during the integration). We represent this unresolved broadening with an effective non-thermal wind dispersion term; it is a pragmatic unresolved-kinematics term---not a direct measure of turbulence or any specific flow geometry. The inclusion of this wind dispersion parameter is further motivated by \citet{lellouch1990io, Lellouch1992, lellouch1996urey} as well as modeling work done by \citet{Gratiy2010}. These works showed that line profiles in the millimeter are too wide to be formed solely by thermal Doppler broadening and that the inclusion of atmospheric dynamics are required to explain the observed line widths. In our implementation, the Doppler broadening line profile function combined the effects of thermal motion and unresolved atmospheric velocity dispersion as

\begin{equation*}
\Delta \nu_D = \frac{\nu_0}{c} \sqrt{\frac{2k_B T}{m} + v^2_{\text{disp}}},
\end{equation*}

\noindent where $\nu_0$ is the rest frequency of the line, T is the local atmospheric temperature, $m$ is the molecular mass, and $v_{\text{disp}}$ the wind dispersion parameter, consistent with line-formation textbooks \citep{Gray_2005}. The thermal and non-thermal components combined to set the total Doppler width of the line profile at each layer in the RT calculation. This combined treatment is appropriate given the low pressures in Io's atmosphere and the dominance of Doppler broadening over collisional effects in this regime. Retrievals performed without $v_{\text{disp}}$ converged to slightly worse goodness of fit and compensated for the missing broadening by inflating the gas temperature and/or the SO$_2$ column density---illustrating the degeneracy and reinforcing the need for the non-thermal dispersion term.

Io's SO$_2$ atmosphere is highly spatially heterogeneous, driven by localized volcanic plumes, unevenly distributed frost deposits, and variable sublimation rates across the surface. This variability implies that emission captured within the ALMA beam may be patchy or partially resolved. If we were to assume a full, uniform beam coverage, any line that appears weaker than predicted by our RT model could be compensated by lowering the SO$_2$ column density and/or temperature---ultimately biasing our retrieved atmospheric parameters. Instead, we explicitly included a fractional beam-filling (fractional coverage) parameter---following the standard treatment adopted in previous millimeter studies of Io's atmosphere. This parameter essentially represents the fraction of the projected surface that is covered by gas at the modeled column and temperature. Including this parameter in our model let us distinguish whether a weaker-than-predicted line is caused by genuinely low SO$_2$ column density/temperature or simply by the gas filling only part of the ALMA beam.

Finally, to produce a model spectrum comparable to our ALMA data, we took the following steps. First, at each frequency we summed the Doppler-shifted and Doppler-broadened brightness temperatures over exactly the same disk-center pixels used in our spectral extraction. \added{We then mimic ALMA's spectral response by convolving the synthetic spectra with an effective instrumental line-spread function, modeled as a symmetric Gaussian with a full width at half maximum equal to the spectral resolution of the ALMA cube. Because the observed spectral lines are broader than the instrumental response (see Figure~\ref{fig:oneline}), reasonable variations in the assumed line-spread function do not materially affect the fitted parameters.} We then interpolated this convolved spectrum onto the observed frequency grid, \added{and a constant (frequency-independent over the narrow fitting window) baseline offset was removed to match the continuum-subtracted data product.} The end result was a synthetic line profile that shared both the spatial aperture and spectral sampling of the measured data. 

For all retrievals presented in this work, we fixed the surface temperature to 115~K with an assumed emissivity of 0.8, consistent with earlier millimeter studies of Io (e.g., \citealt{deKleer2024}; \citealt{dePater2020b}). To test the sensitivity of this assumption, we performed a dedicated set of retrievals on the M22-L dataset where we varied the fixed surface temperature between 100, 115, and 130~K. As detailed in Appendix~\ref{sec:AppendixSurfaceTemp}, fixing the surface temperature had little qualitative impact on the shape of the retrieved atmospheric temperature profile, \added{even though other correlated parameters adjusted}. Because temperature and SO$_2$ column density---as well as the other retrieved parameters---are partially degenerate, higher assumed surface temperatures pushed the best-fit SO$_2$ column density upward. We constrained this trade-off between parameters and gained vertical leverage (sensitivity) on the atmosphere by fitting SO$_2$ lines that span a range of opacities and lower-state energies. Optically thin lines are roughly column-weighted and thus emphasize deeper, higher-pressure layers. For optically thick lines, the line core forms at higher altitudes (lower pressure) while the wings sample deeper (higher pressure) gas. In our best-fit models, the disk-center, line-center optical depths for the SO$_2$ transitions span $\tau_{\nu_0} \approx 0.1 - 30$. The weakest lines have $\tau_{\nu_0} \lesssim 1$ and are optically thin to moderately thick, whereas the strongest lines reach $\tau_{\nu_0} \sim 20-30$ and are in the optically thick regime. In the joint fit, one temperature profile must reproduce these line shapes and relative strengths across all lines at once, which curbed degeneracies among parameters. In our modeling, the atmospheric temperature set the excitation and thermal share of the line width; SO$_2$ column raised opacity (scaling thinner lines and, once saturated, broadening the apparent profile) without changing the intrinsic Doppler width; $v_{\text{disp}}$ added nearly uniform non-thermal broadening; and fractional coverage mainly scaled amplitudes. To keep the inference physically plausible and mitigate degeneracies, we applied physically-based priors on our retrieved parameters, as we discuss later in Subsection~\ref{subsec: Priors}. With these constraints, the remaining parameters correlations were modest and the retrieved TP shape was stable across the 100-130~K surface-temperature tests (Appendix~\ref{sec:AppendixSurfaceTemp}). \added{We therefore concluded that a 115~K fixed surface temperature is an appropriate choice and introduced no significant bias in the inferred TP profile shape.}

\subsection{Bayesian Inference and MCMC Sampling}\label{subsec: MCMC}
The forward model described above depends on a set of physical parameters that jointly determine the emergent disk-center spectra. In all our retrievals, we treated the SO$_2$ column density, the temperatures at each TP node, the transition pressure (for the 3T models), the fractional beam-filling factor, and the sub-beam wind-dispersion velocity as freely sampled physical parameters. In addition to these atmospheric parameters, the statistical components of the likelihood (the dataset-level jitter noise term and the degrees-of-freedom parameter of the Student's-t residual model) were also sampled as part of the full parameter vector, and are introduced in the subsections that follow. To explore how all of these parameters interact and to derive robust uncertainty estimates, we adopted a Bayesian framework.

We implemented our Bayesian retrieval in Python using the affine invariant MCMC ensemble sampler, \texttt{emcee} \citep{emcee}. All retrieved physical model parameters were fit simultaneously; these include the SO$_2$ column density, the temperatures at each of the TP nodes, the transition pressure defining the three‐node profile, the beam-filling fraction, and the sub-beam wind-dispersion velocity. Walkers \added{(the individual MCMC chains that sample the parameter space in parallel) }were initialized in a Gaussian spread around an isothermal starting profile and each temperature node, regardless of model complexity, was initially set to 250 K---a representative isotherm within the range inferred from prior mm/sub-mm retrievals \citep{Lellouch1992, Moullet2010, Roth2020, dePater2020b, deKleer2024}. We employed 50 walkers (twice the number of CPU cores used) and ran an exploratory chain for 2,000 steps. After this burn-in, exploratory phase, we re-centered the walkers on their median positions and proceeded with up to a 50,000 step production run. However, we often halted the run early once convergence was confirmed by both visual inspection of trace and corner plots and/or quantitative \texttt{ArviZ} \citep{arviz_2019} diagnostics. These diagnostics included the Gelman-Rubin $\hat{R}$ convergence statistic (potential scale reduction factor; values near 1 indicate that independent chains have mixed well) staying below $\sim$1.01 and effective sample sizes exceeding 400 (the number of independent draws after accounting for autocorrelation; \citealt{Vehtari2021}). We also monitored the mean acceptance fraction, keeping it between 20$\%$ and 50$\%$. Corner plots for the three-node (3T) temperature profile applied to the leading hemisphere data and for the one-node (1T) profile applied to the trailing hemisphere data\added{---reflecting the differing vertical sensitivity of the two hemispheres discussed in Section~\ref{sec:Discussion}---are available in a Zenodo repository (DOI: \href{https://doi.org/10.5281/zenodo.18342667}{10.5281/zenodo.18342667}).}

At each MCMC step, the full parameter set drove our RT routine to generate disk-centered spectra for every targeted line. These modeled transitions were then interpolated onto the observed frequency grid. For every channel we define the residual,
\begin{equation*}
r_i = F^{\text{obs}}_i - F^{\text{mod}}_i ,
\end{equation*}
where i indexes the frequency channels, $F^{\text{obs}}_i$ is the observed continuum-subtracted flux, and $F^{\text{mod}}_i$ is the corresponding model prediction. We modeled the residuals with a likelihood based on the Student's t-distribution, 
\begin{equation*}
p\!\left(r_i \mid \sigma_{\text{tot}}, \nu\right)=
\frac{\Gamma\!\bigl(\tfrac{\nu+1}{2}\bigr)}
     {\Gamma\!\bigl(\tfrac{\nu}{2}\bigr)\sqrt{\nu\pi}\,\sigma_{\text{tot}}}
\Bigl[1+\tfrac{r_i^{2}}{\nu\,\sigma_{\text{tot}}^{2}}\Bigr]^{-(\nu+1)/2},
\label{eq:student_t}
\end{equation*}
where $\nu$ is the degrees-of-freedom parameter controlling the heaviness of the Student-t tails (large $\nu \implies$ Gaussian distribution), $\Gamma(X)$ is the Euler gamma function, and $\sigma_{\text{tot}}$ is the noise parameter. We chose this distribution for its built-in robustness to the occasional channel-level outliers that can survive ALMA flagging and calibration. We take
\begin{equation*}
\sigma_{\text{tot}}
   =\sqrt{\sigma_{\text{thermal}}^{2}+\sigma_{\text{jitter}}^{2}},
\end{equation*}
where $\sigma_{\text{thermal}}$ is the per-line thermal noise estimated from line-free channels (as described in Section~\ref{sec:Observations and Data Reduction}), and $\sigma_{\text{jitter}}$ is a variance-inflation term that can slightly raise the effective noise floor to prevent small amplitude mismatches from dominating the likelihood. Both $\sigma_{\text{jitter}}$ and $\nu$ were fit concurrently with the physical model parameters, and the inclusion of the Student's t-distribution and $\sigma_{\text{jitter}}$ was important for retrieving stable and reliable model comparison results. We constructed the joint likelihood by summing the individual log-likelihoods of all targeted transitions, which naturally weighted each line according to its signal-to-noise. We ran the sampler in parallel across CPU cores, combined with in-memory caching of spectra, to ensure efficient sampling. 

After completing the MCMC sampling and verifying convergence, we extracted the median parameter vector from the posterior distribution and used it to initialize a deterministic optimization routine. Specifically, we applied the Nelder-Mead minimization algorithm as implemented in the \texttt{scipy.optimize} module of the \texttt{SciPy} library \citep{virtanen2020scipy}. \added{This procedure maximized the same log posterior used in the sampler, yielding a maximum-a-posteriori parameter vector.} The resulting best-fit model is used for all diagnostic plots (e.g., spectral fits and temperature profiles) shown throughout the paper. All quoted 1$\sigma$ uncertainties were derived from the posterior by propagating random parameter draws through the forward model and taking the resulting 16th-84th percentile range at each layer (for TP profiles) or channel (for spectra). 

\subsection{MCMC Priors}\label{subsec: Priors}
\subsubsection{Priors on Noise and Hyper-parameters}
For the dataset-level jitter, $\sigma_{\text{jitter}}$, we adopted a Half-normal prior centered at zero with a scale equal to 20$\%$ of the median absolute flux across all fitted spectra for Band 8 datasets. This conservative choice is motivated by the ALMA Technical Handbook, which reports typical Band 8 relative flux calibration uncertainties of $\sim\!10\%$ (1$\sigma$) and notes that additional flux-transfer errors can---in unfavorable cases---double this, implying effective uncertainties up to $20\%$ \citep{ALMA_THB}. Our $20\%$ prior scale for Band 8 is therefore intended to encompass this upper end of the plausible calibration range. For Band 7 datasets, we adopted a $10\%$ prior width which is consistent with the stated accuracy reported in the handbook \citep{ALMA_THB}.

For the degrees-of-freedom, $\nu$, which controls the heaviness of the Student's t-distribution tails, we adopted an inverse gamma prior, $\nu \sim \text{Inv-}\Gamma(4, 15)$, following the recommendation of \citet{martin2024robust}. While this prior is originally smooth and proper over $\nu > 0$, we truncated it to enforce $\nu > 2$ \citep{ivezic2020statistics}. This choice ensured that the likelihood retains a well-defined mean and variance, which is important for maintaining stability in parameter estimation and uncertainty quantification. Although this truncation introduced a sharp cutoff, it preserved the desirable shape of the \citet{martin2024robust} prior over the physically relevant regime. Additionally, in our retrieval for most cases, the posterior peaked above $\nu = 2$, and the walkers did not clump at the boundary---indicating that the results were not sensitive to the imposed cutoff. In the few fits where samples accumulated near $\nu = 2$, refitting with $\nu$ fixed at 5 yielded similar results, indicating that our results are insensitive to the exact tail weight. We therefore retained the truncated prior without further adjustment.

\subsubsection{Priors on the SO$_2$ Column Density}
Independently constraining Io's temperature profile and SO$_2$ column density can lead MCMC walkers into regions of parameter space that---while allowed by minor improvements in the likelihood---correspond to nonphysical combinations of low temperature and high column or vice versa. In exploratory retrievals performed without any prior on the column, or with very broad uniform priors, the sampler typically appeared stable for several thousand steps before drifting into these clearly nonphysical low-$T$/high-$N$ and high-$T$/low-$N$ regimes, particularly for the 3T model where the additional flexibility amplifies the underlying degeneracy. To aid our retrieval, we adopted Gaussian priors on the base-10 logarithm of the SO$_2$ column density separately for Io's leading and trailing hemispheres. We motivated these priors based on the long-term IRTF/TEXES survey of \citet{Giles2024}, which spanned 22 years of equatorial SO$_2$ column density measurements between 2001 and 2023. 

In this survey, \citet{Giles2024} determine how Io's SO$_2$ column density varied with both season and longitude by fitting mid-infrared absorption lines across nearly two Jovian years. From the longitude SO$_2$ column density profile of \citet[][Fig.~2]{Giles2024} we took $N_{180^{\circ}}=1.8\times10^{17}$, $N_{90^{\circ}}=1.0\times10^{17}$, and $N_{270^{\circ}}=0.6\times10^{17}\ \mathrm{cm^{-2}}$, giving scale factors of $f_{leading} \approx 0.6$ (90$^{\circ}$ W) and $f_{trailing} \approx 0.3$ (270$^{\circ}$ W) relative to the anti-Jovian value at 180$^{\circ}$ W. These factors were required because \citet{Giles2024} provides time-resolved SO$_2$ columns only at the anti- and sub-Jovian longitudes; we applied these scaling factors to translate the anti-Jovian measurements to the 90$^{\circ}$ W and 270$^{\circ}$ W longitudes sampled in our ALMA data. From the seasonal anti-Jovian record \citet[][Fig.~3]{Giles2024} we took $N_{180^{\circ}}=2.0\times10^{17}$~cm$^{-2}$ for our M22-L/T datasets and $1.5\times10^{17}$~cm$^{-2}$ for our J21-L/T datasets, then scaled them by the scale factors above for each hemisphere to obtain the hemisphere-specific means listed in Table \ref{tab:priors}. We set the prior width to $\sigma = 0.75$ in log$_{10}$N (a factor of $\sim\!5.6$ in linear column density). This width left room for revision of earlier isothermal estimates and accommodates transient volcanic surges in SO$_2$ column density. Therefore, the prior was loose enough for the ALMA data to drive the posterior, yet still tight enough to restrict the clearly nonphysical low-$T$/high-$N$ and high-$T$/low-$N$ extremes, and its role is stabilizing the temperature retrieval rather than imposing any particular SO$_2$ value.

\begin{deluxetable*}{lcccc}
\tablecaption{Gaussian priors on the SO$_2$ column density
\label{tab:priors}}
\tablehead{
\colhead{Epoch} &
\colhead{Hemisphere} &
\colhead{$\log_{10}(N)$} &
\colhead{Central value $N$ [cm$^{-2}$]} &
\colhead{$1\sigma$ interval [cm$^{-2}$]}
}
\startdata
2021 July & Leading  & 16.95 & $8.9\times10^{16}$ & $(1.6$-$50)\times10^{16}$ \\
         & Trailing & 16.65 & $4.5\times10^{16}$ & $(0.79$-$25)\times10^{16}$ \\[2pt]
2022 May & Leading  & 17.08 & $1.2\times10^{17}$ & $(0.21$-$6.8)\times10^{17}$ \\
         & Trailing & 16.78 & $6.0\times10^{16}$ & $(0.11$-$3.4)\times10^{17}$ \\
\enddata
\tablecomments{Each prior is Gaussian in $\log_{10}(N)$ with a width of $0.75$ in logarithmic space; this corresponds to multiplying or dividing the central value by a factor of $10^{0.75}\!\approx\!5.6$ to obtain the $\pm1\sigma$ limits in linear space.}
\end{deluxetable*}

\subsubsection{Prior on the Frost-Equilibrium Constraint}
In addition to the SO$_2$ column prior, we enforced a frost-equilibrium constraint on the surface pressure because SO$_2$ frost sublimation helps sustain Io's atmosphere \citep{tsang2013sublimation}. If the surface pressure were too high, the frost-equilibrium temperature could exceed the actual surface temperature, causing net SO$_2$ deposition instead of sublimation---effectively collapsing the atmosphere. To prevent this non-physical atmospheric state, we inverted \citet{wagman1979sublimation}'s empirical vapor-pressure law ($P_{vapor} = 1.52\times10^{8}e^{-4510/T}$) for each sampled surface pressure to compute the corresponding frost point. Since Io's atmosphere is patchy and time-variable, partly buffered by non-condensable species and volcanic outgassing, and because our single-beam, disk-center spectrum likely averaged over sub-beam departures from strict equilibrium, we imposed a soft (one-sided Gaussian) penalty---not a hard cutoff---whenever the computed frost point exceeded the assumed, fixed surface temperature of 115 K used in our main retrievals. In the sensitivity tests described in Appendix~\ref{sec:AppendixSurfaceTemp}, the same penalty was applied relative to the corresponding test surface temperature rather than the nominal 115~K. This constraint kept the MCMC focused on pressure-temperature combinations that can physically sustain frost sublimation, while still allowing the ALMA data to drive the fit.

We adopted a 1$\sigma$ width of 5~K for this frost-equilibrium penalty. This choice was informed by the extreme sensitivity of SO$_2$ vapor pressure to temperature as even small changes of a few kelvin can dramatically alter equilibrium pressure by tens to over one hundred percent at typical Io temperatures. Additionally, this 5~K width allowed our model to account for the observed complex range of atmospheric behavior, as described below. 

While pure SO$_2$ sublimation models predict atmospheric pressure drops by orders of magnitude in eclipse, observations show only a factor-of-few decrease \citep{tsang2016collapse, dePater2020b}. Mid-infrared eclipse measurements by \citet{tsang2016collapse} detected a disk-integrated SO$_2$ column density dropping by a factor of $5 \pm 2$ from pre-eclipse to mid-eclipse. However, Io's surface temperatures drop significantly from day to night (e.g., 127~K to 105~K as observed by \citealt{tsang2016collapse}), which under pure vapor equilibrium would imply a much larger predicted pressure change than observed. This mismatch between the large pressure change predicted by vapor-pressure equilibrium models and the much smaller change actually observed suggests that Io's atmosphere is not fully governed by pure frost equilibrium alone. Instead, it must be buffered by additional processes such as volcanic supply, non-condensable species, or spatially heterogeneous sublimation \citep{Moore2009, dePater2020b}. 

We can quantify this atmospheric damping by using the temperature sensitivity of the SO$_2$ vapor pressure from Wagman's law: $d(\ln P)/dT = 4510/T^2$. At temperatures relevant to SO$_2$ sublimation on Io (e.g., around 115-120 K), this relationship demonstrates that a change of approximately 5~K in temperature is sufficient to cause a factor-of-five change in SO$_2$ equilibrium vapor pressure. This directly links the observed factor-of-five column density change in \citet{tsang2016collapse} to an effective $\sim$5~K shift in frost-equilibrium temperature for the SO$_2$ gas itself. Since this $\sim$5~K shift matches the scale implied by the observations, a 1$\sigma$ width of 5~K naturally encodes the observed atmospheric variability. 

Millimeter-wave ALMA observations \citep{dePater2020b} provide additional, independent support for our adopted 5~K frost-equilibrium prior by revealing spatially resolved changes in Io's atmosphere during eclipse, consistent with a thermally buffered and spatially selective collapse. \citet{dePater2020b} found that the disk-integrated SO$_2$ column densities remained approximately constant before and after eclipse, while the fractional coverage decreased by a factor of 2-3. This suggests that the atmospheric collapse during eclipse may involve spatial retreat rather than a global pressure drop (as suggested by the factor of $5\pm2$ drop in disk-integrated SO$_2$ column density measured by \citet{tsang2016collapse}). The apparent discrepancy is largely explained by the difference in spatial sensitivity: \citet{dePater2020b} was able to disentangle column density from areal coverage, resolving regions with high column but limited extent, whereas \citet{tsang2016collapse} could only measure disk-averaged totals and thus could not distinguish between a thinner global atmosphere and a thicker, patchier one. Both studies nevertheless indicate that a substantial portion of Io's atmosphere disappears during eclipse, yet not to the extent predicted by pure vapor-pressure equilibrium. Moreover, the spatial distribution reported by \citet{dePater2020b} suggests that the sublimation-supported component collapses away in eclipse, while the remaining SO$_2$ is sustained either by direct volcanic outgassing or by localized frost sublimation triggered at advancing lava flows. The residual atmosphere is likely further buffered by non-condensable species, such as SO, which remain in the gas phase under typical eclipse temperatures and provide a background pressure that buffers the atmosphere against a complete freeze-out \citep{Moore2009, dePater2020b}. Together, these results highlight that Io's atmosphere is neither a simple equilibrium veneer nor fully volcanic in origin, but it is a hybrid system in which multiple, complex processes govern its stability.

By setting a 5~K $\sigma$, our model was able to capture the observed scale of atmospheric damping and stability across different wavelengths, acknowledging both the significant disk-integrated changes implied by \citet{tsang2016collapse} and the more stable local conditions observed by \citet{dePater2020b}. This provided a robust, yet soft, constraint that prevented the MCMC from exploring nonphysical SO$_2$ condensation states, while remaining flexible enough to account for the complex nature of Io's atmosphere and allowing the ALMA data, with its implications for local stability and fractional coverage, to drive the fit.

\subsubsection{Physical Bounds on Atmospheric Parameters}
In addition to the aforementioned Gaussian priors, our MCMC retrieval imposed several hard physical bounds on the atmospheric parameters to ensure that sampled solutions remained physically realistic. These included requiring all atmospheric temperatures to remain above 80~K (aligning with the lower end of reported temperatures discussed in Section~\ref{subsec: Temp-limitations}) and below 4,000~K (well above even the hottest temperatures predicted in the very low pressure layers). Fractional surface coverage parameters (e.g., $f_c$) were restricted to the range [0, 1], and wind dispersion parameters were constrained between 0 and 1,000 $\mathrm{m~s^{-1}}$. These bounds prevented the MCMC from exploring regions of parameter space that were clearly nonphysical, thereby improving computational efficiency and the reliability of the retrieved posteriors.

Above temperatures of 1,800~K, we additionally applied a soft Gaussian penalty with a width of 300~K. This was not intended to impose a physical upper limit on Io's atmosphere. Instead, this penalty addressed a numerical degeneracy: at the lower pressures probed by the upper-node (in the 2T and 3T models), very high temperatures alter the Doppler width only weakly in the emergent spectrum. Changes from $\sim\!1,000$~K to several thousand kelvin produce differences far below the noise level and can be absorbed by small adjustments in column and/or vertical structure---leaving the likelihood essentially flat in that direction. 
\added{Without this soft Gaussian penalty, a handful of walkers can drift toward arbitrarily high upper-node temperatures without any corresponding change in likelihood, despite the best-fit temperature profile and the structure at lower pressures remaining stable. This penalty therefore does not change the inferred temperature profile in the data-sensitive region, but simply prevents uninformative drift and allows finite, well-defined credible intervals to be reported for the upper-node.}

\subsection{Model Comparison via PSIS-LOO-CV}\label{subsec: PSIS}
\added{After verifying convergence and sufficient mixing of our walkers, we discarded the burn-in phase and applied thinning as needed. At each MCMC step, we computed and stored the per-channel log-likelihood contributions across all fitted SO$_2$ lines. These pointwise likelihood vectors were stored alongside the posterior samples for subsequent model comparison.} We evaluated relative model performance using Pareto-smoothed importance sampling leave-one-out cross-validation (PSIS-LOO-CV; \citealt{vehtari2024paretosmoothedimportancesampling, Vehtari_2016}). \added{PSIS-LOO-CV uses importance sampling to estimate how well a model predicts each spectral channel when that channel is left out from the fit. This approach penalizes overfitting because overly flexible models can adjust their global parameters to accommodate noise in specific channels, thereby improving the full-data likelihood but yielding poorer predictive scores once those channels are excluded.} The resulting metric of PSIS-LOO-CV, the expected log pointwise predictive density $\widehat{\mathrm{elpd}}_{\mathrm{loo}}$, thus \added{implicitly} balances goodness-of-fit with model complexity. For ease of interpretation, we report a leave-one-out-information-criterion, $\mathrm{LOOIC} = -2\times\,\widehat{\mathrm{elpd}}_{\mathrm{loo}}$, so that lower values indicate better expected performance for left-out data points of the same type. This is similar in spirit to the Akaike (AIC) and Deviance (DIC) information criteria. We adopt LOOIC as our information criterion since it directly estimates predictive accuracy from the posterior without making underlying assumptions about the posterior shape. Additionally, through this approach, we obtained a direct estimate of the uncertainty on $\widehat{\mathrm{elpd}}_{\mathrm{loo}}$, which reflects the sampling variability in the model's predictive performance. This uncertainty provides a principled way to assess whether observed differences in LOOIC between models are statistically meaningful. In practice, we interpreted model differences as such: when the absolute $\Delta\mathrm{LOOIC}$ exceeded the combined (in quadrature) standard error on that difference, we considered the preference as statistically meaningful. When this threshold was not met, we considered the models as statistically indistinguishable in their predictive performance. In these situations, physically plausible models with comparably small LOOIC values still warrant discussion because they represent alternative temperature structures that the data do not meaningfully distinguish.

\section{Results}
\label{sec:Results}
From our four datasets, we retrieved Io's vertical temperature profile, SO$_2$ column density, wind dispersion, fractional coverage, and transition pressure (when applicable) using the SO$_2$ rotational lines tabulated in Table~\ref{tab:so2_lines}. In this Section, we present results from all four datasets. For conciseness, the main text shows the retrieved temperature profiles and the corresponding contribution functions and spectral line fits are provided in Appendix~\ref{sec:AppendixContrib} and Appendix~\ref{sec:AppendixFigures}, respectively.

\subsection{Leading Hemisphere}
\begin{figure*}
    \centering
    \includegraphics[width=\textwidth]{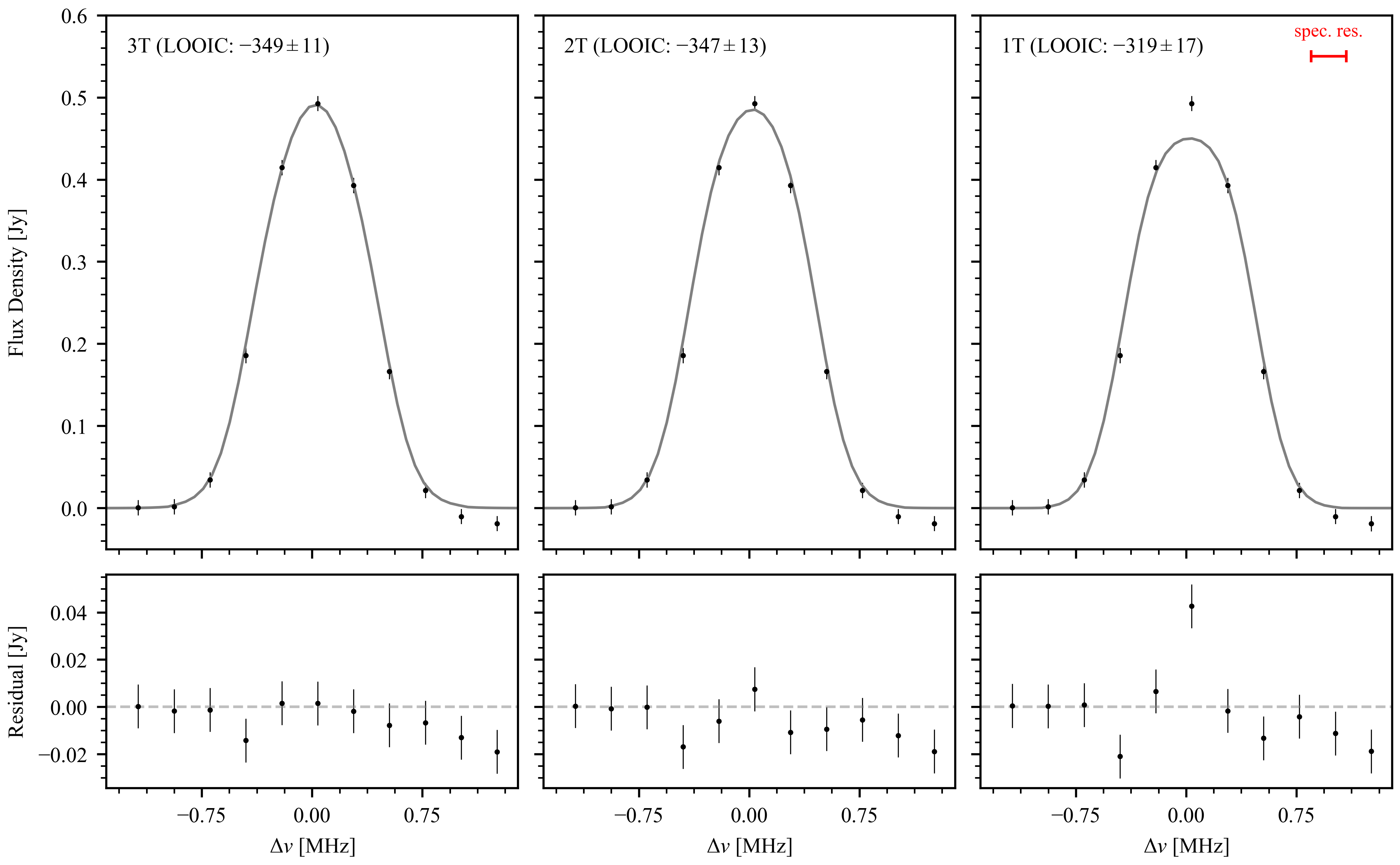}
    \caption{\textbf{Model Comparison for the SO$_2$ transition at 430.194 GHz from M22-L.} Observed spectrum (black points with error bars) overlaid with best-fit model spectra for the three tested temperature profiles: 3T (left), 2T (middle), and 1T (right). \added{The red bar at the top right of the third column indicates the ALMA spectral resolution (FWHM).} Residuals are shown in the lower row.}
    \label{fig:oneline}
\end{figure*}

For the leading hemisphere dataset, M22-L, we found that, among the tested temperature parameterizations described in Section~\ref{sec:Modeling Framework}, the three-node (3T; $\mathrm{LOOIC} = -349 \pm 11$) and two-node (2T; $\mathrm{LOOIC} = -347 \pm 13$) profile yielded statistically comparable fits, with their LOOIC values differing by only $\Delta\mathrm{LOOIC} = 2$, which was well within the combined standard error. In contrast, the one-node, isothermal profile (1T; $\mathrm{LOOIC} = -319 \pm 17$), was disfavored by $\Delta\mathrm{LOOIC}\approx 30$ relative to the 3T model and by $\Delta\mathrm{LOOIC} = 28$ relative to the 2T model. Both differences exceeded the respective combined standard errors, indicating statistical support for the added flexibility of the multi-node parameterizations. Among the SO$_2$ transitions we fit for this dataset, we highlight the transition at $\nu_{0} = 430.194$ GHz because it is the strongest transition in the M22-L set (Table~\ref{tab:so2_lines}) and has the highest signal-to-noise, which makes differences between the 1T and multi-node models most clearly visible. In Figure~\ref{fig:oneline}, we demonstrate this improvement, and show that the 3T and 2T models both match the overall line profile for this transition. Meanwhile, the 1T model shows its largest deviation at the central data point, where it underestimates the observed flux density. Smaller-scale deviations were present in the other transitions, but these were generally within the observational uncertainties. 

\begin{figure*}
    \centering
    \includegraphics[width=\textwidth]{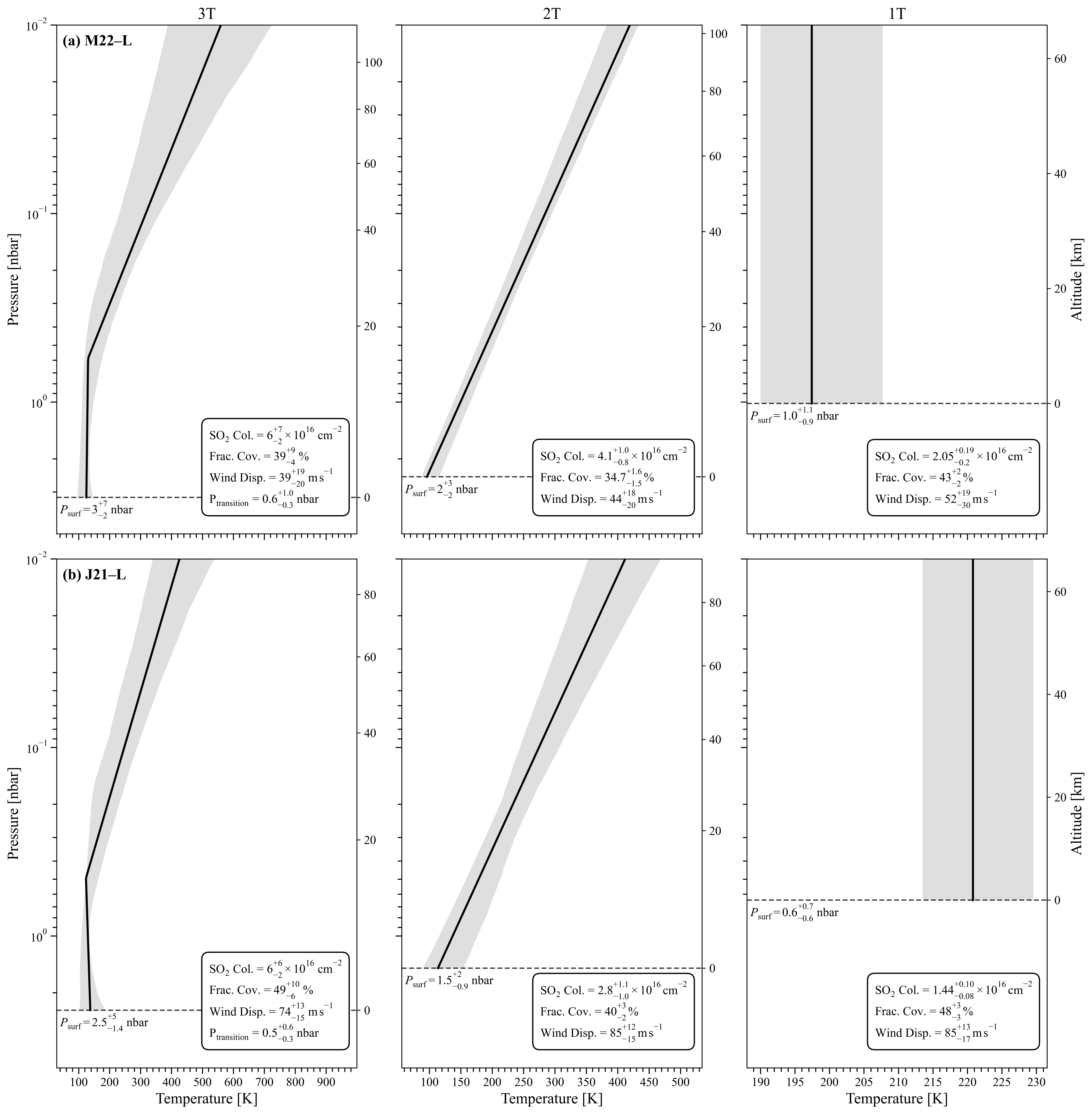}
    \caption{\textbf{Retrieved temperature profiles for leading hemisphere datasets: M22-L (a, top row) and J21-L (b, bottom row).} Columns show the 3-node (3T), 2-node (2T), and 1-node (1T) parameterizations. The black curve is the best-fit TP; the gray band marks the 1$\sigma$ credible interval. The legend for each profile provides the best-fit SO$_2$ column density, fractional coverage, wind dispersion parameters, and, when applicable, the transition pressure (with $1\sigma$ errors). Left axes give pressure; right axes show the corresponding altitude. The dashed horizontal black line indicates the modeled surface pressure, quoted with uncertainties calculated from the upper and lower bounds of the SO$_2$ column density.}
    \label{fig: leading-temp-profiles}
\end{figure*}

For our J21-L leading hemisphere dataset, we found that the 3T model ($\mathrm{LOOIC} = -519\pm28$) yielded the minimum LOOIC value. However, the differences relative to the 1T ($\mathrm{LOOIC} = -506\pm28$) and 2T ($\mathrm{LOOIC} = -505\pm28$) models were smaller than the standard errors, indicating that the statistical significance for the 3T model was weak in this dataset. The retrieved 1T, 2T, and 3T temperature profiles, along with best-fit values for the SO$_2$ column density, fractional coverage, wind dispersion, and, when applicable, the transition pressure, are shown for both leading hemisphere datasets in Figure~\ref{fig: leading-temp-profiles}. We note that the 3T profile retrieved for the J21-L dataset qualitatively matches the 3T structure found in the better constrained, M22-L dataset, supporting its physical plausibility. 

In Appendix~\ref{sec:AppendixContrib}, we show the line contribution functions corresponding to the retrieved temperature profiles and reason why the 3T profile yielded vertical leverage. In short, 1T fitted a single isotherm and a relatively low surface pressure, which confined the contribution-function peaks to a narrow pressure range and truncated several of them at the boundary; 2T added a single gradient that shifted the bulk of those peaks together without strongly increasing their spread in pressure; 3T permitted a cool lower layer and warmer upper layer, which separated the peaks by pressure so high-altitude lines can brighten without distorting the fit of the deeper-forming lines. This same separation also reduced covariance between retrieved parameters, since upper-layer emission can be matched by the upper-layer temperature rather than by changes in other atmospheric parameters. For reasons motivated in Section~\ref{subsec: why3T}, we use the 3T profiles for our physical interpretation of Io's atmosphere. The modest LOOIC differences relative to 2T (and, in J21-L, to 1T) indicate that this choice is pragmatic rather than a uniquely preferred structure. The best-fit 3T model fit across our targeted SO$_2$ transitions are shown for the M22-L and J21-L dataset in Appendix~\ref{sec:AppendixFigures} (Figure~\ref{fig:stacked_spectra}).

\subsection{Trailing Hemisphere}
For the trailing hemisphere dataset, M22-T, we found that the 1T ($\mathrm{LOOIC} = -529 \pm 13$), 2T ($\mathrm{LOOIC} = -528 \pm 12$), and 3T ($\mathrm{LOOIC} = -527 \pm 12$) temperature profiles yielded statistically indistinguishable fits. The LOOIC differences were well within their standard errors and thus there was no strong preference for additional temperature nodes in this trailing hemisphere dataset. 

Similarly, for the J21-T trailing hemisphere dataset, the 1T ($\mathrm{LOOIC} = -604 \pm 18$), 2T ($\mathrm{LOOIC} = -602 \pm 17$), and 3T ($\mathrm{LOOIC} = -601 \pm 17$) models differed only by $\Delta\mathrm{LOOIC} \approx 1-3$, again far below the combined uncertainty. As with M22-T, these results indicate that the simpler isothermal parameterization performed just as well, in a predictive sense, as the more flexible multi-node temperature profiles for the trailing hemisphere.

\begin{figure*}
    \centering
    \includegraphics[width=\textwidth]{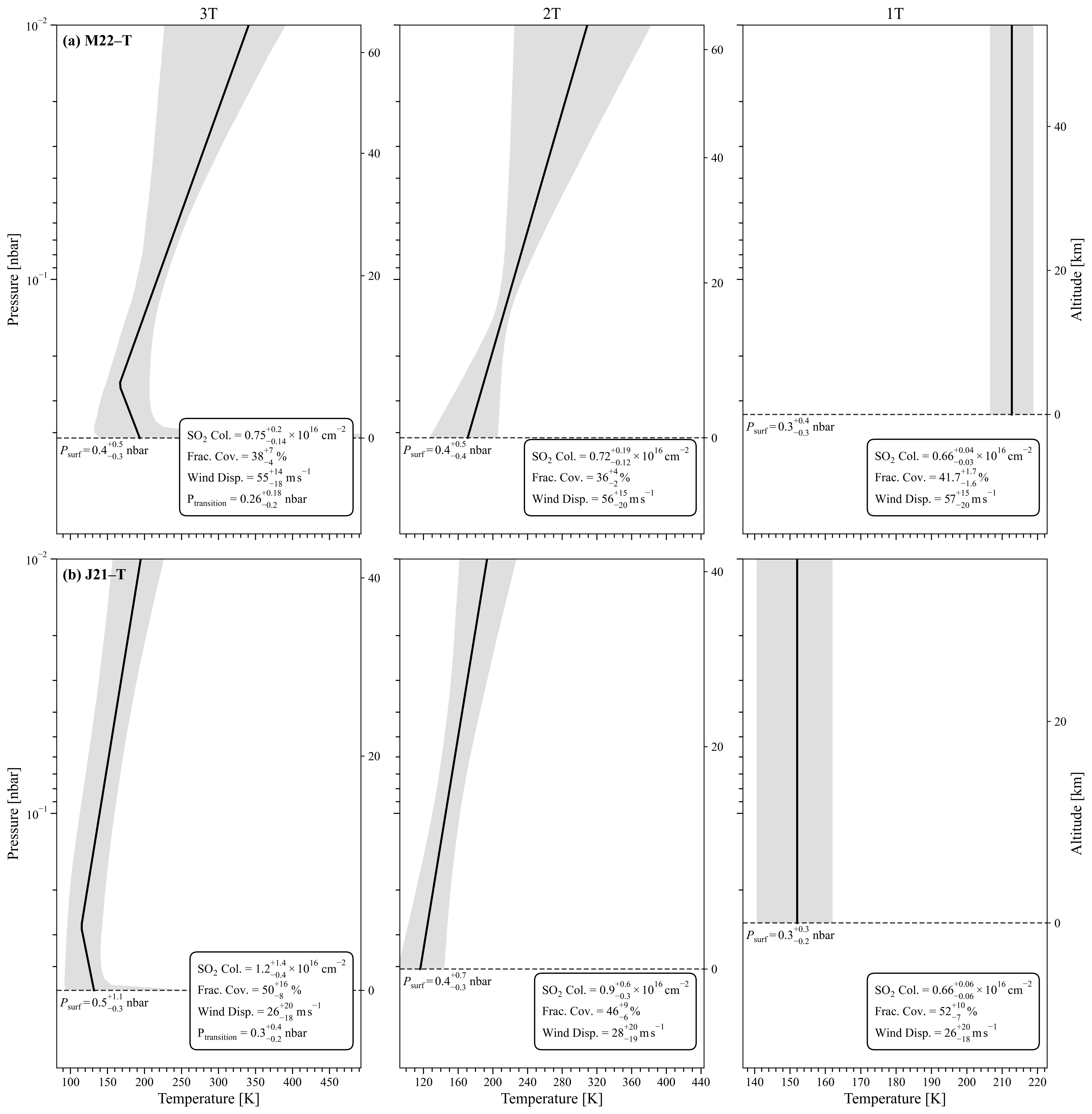}
    \caption{\textbf{Retrieved temperature profiles for trailing hemisphere datasets: M22-T (a, top row) and J21-T (b, bottom row).} Columns show the 3-node (3T), 2-node (2T), and 1-node (1T) parameterizations. The black curve is the best-fit TP; the gray band marks the 1$\sigma$ credible interval. The legend for each profile provides the best-fit SO$_2$ column density, fractional coverage, wind dispersion parameters, and, when applicable, the transition pressure (with $1\sigma$ errors). Left axes give pressure; right axes show the corresponding altitude. The dashed horizontal black line indicates the modeled surface pressure, quoted with uncertainties calculated from the upper and lower bounds of the SO$_2$ column density.}
    \label{fig: trailing-temp-profiles}
\end{figure*}

The retrieved 1T, 2T, and 3T temperature profiles, along with best-fit values for the SO$_2$ column density, fractional coverage, wind dispersion, and, when applicable, the transition pressure, are shown for both trailing hemisphere datasets in Figure~\ref{fig: trailing-temp-profiles}. Across hemispheres, we found most retrieved parameters to be comparable; however, the trailing side SO$_2$ column was consistently $\sim3-10$ times lower than those retrieved on the leading side. This lower column pushed optical depth unity close to the surface and clustered the contribution functions at low altitudes, reducing vertical leverage and signal-to-noise. This column-driven, low-altitude weighting explains the lack of model preference and is further discussed in Appendix~\ref{sec:AppendixContrib}, where we also show the line contribution functions corresponding to the retrieved temperature profiles. Unlike for the leading hemisphere, we do not single out a statistically preferred TP model for the trailing side, since the LOOIC values are indistinguishable across 1T, 2T, and 3T. However, we do note that the trailing 2T and 3T retrievals yielded qualitatively similar profiles to those retrieved on the leading hemisphere, with a relatively cold lower atmosphere and temperatures that increase toward lower pressures above it. The best-fit 1T model fit across our targeted SO$_2$ transitions are shown for the M22-T and J21-T dataset in Appendix~\ref{sec:AppendixFigures} (Figure~\ref{fig:stacked_spectra}). We show the fits from the 1T retrieval because the 2T and 3T models produce identical spectral fits.

\subsection{Summarized Results}
Across the four datasets, only the Band 8 leading hemisphere spectra (M22-L) showed statistical evidence in our LOOIC comparison against an isothermal atmosphere. In that dataset, the 3T and 2T profiles have comparable LOOIC values and both outperform the 1T model---indicating that vertical temperature structure is statistically favored, although not uniquely determined. The Band 7 leading hemisphere dataset (J21-L) has weaker model discrimination, but the retrieved 2T and 3T temperature structures closely match those found for M22-L. For the 3T model, the physical interpretation of this structure, including its connection to previously proposed mesopause-like temperature minima, is discussed in the following section.

In contrast to the leading hemisphere, both trailing hemisphere datasets (M22-T, J21-T) showed no significant differences in predictive performance among the 1T, 2T, and 3T models. For the trailing side, additional temperature nodes were not uniquely required by the data. However, we note that the retrieved multi-node temperature profiles for the trailing side are qualitatively consistent with the retrieved multi-node profiles from the leading side. For all four datasets, the contribution functions in Appendix~\ref{sec:AppendixContrib} show that our spectra were mainly sensitive to pressures greater than $\sim$10$^{-2}$~nbar; therefore, we limit our interpretation to temperature retrieved at pressures greater than $\sim$10$^{-2}$~nbar. 

Retrieved SO$_2$ column densities (and hence the surface pressure) were systematically higher on the leading hemisphere than on the trailing hemisphere, with fractional coverage values typically in the $35 - 50\%$ range, for both hemispheres. Wind dispersions spanned $\sim\!26-85 \mathrm{~m~s^{-1}}$ across datasets, which we interpret as additional LOS velocity dispersions on top of the thermal broadening. These values are comparable to, though somewhat smaller than, the tens to few-hundred $\mathrm{m~s^{-1}}$ SO$_2$ winds inferred from Doppler mapping \citep{moullet2008first, Thelen2025}, and are therefore physically reasonable for unresolved atmospheric motions within the beam.

\added{Table~\ref{tab:summary_results} summarizes these retrieval results, listing the best-fit values with 68$\%$ credible intervals for each atmospheric parameter as well as the LOOIC scores and their standard errors for each model and dataset. The retrieved temperature profiles---with their corresponding 1$\sigma$ credible bounds---are shown in Figures~\ref{fig: leading-temp-profiles} and ~\ref{fig: trailing-temp-profiles}.}

\begin{deluxetable*}{ccccccccccc}
\tabletypesize{\footnotesize}
\tablewidth{0pt}
\tablecaption{Model Comparison and Retrieved Atmospheric Parameters\label{tab:summary_results}}
\tablehead{
\colhead{Dataset} &
\colhead{Model} &
\colhead{$N_{\mathrm{SO_2}}$} &
\colhead{$P_{\rm surf}$} &
\colhead{$T_1$} &
\colhead{$T_2$} &
\colhead{$T_3$} &
\colhead{$P_{\mathrm{transition}}$} &
\colhead{Frac. Cov.} &
\colhead{Wind Disp.} &
\colhead{LOOIC}
\\[-3pt]
\colhead{} & \colhead{} &
\colhead{[$10^{16}\,\mathrm{cm^{-2}}$]} &
\colhead{[nbar]} &
\colhead{[K]} &
\colhead{[K]} &
\colhead{[K]} &
\colhead{[nbar]} &
\colhead{[\%]} &
\colhead{[m\,s$^{-1}$]} &
\colhead{[$\pm$ SE]}
}
\startdata
\multicolumn{11}{c}{\text{Leading Hemisphere}} \\
\tableline
\noalign{\vskip 0.6ex}
        & 3T & $6^{+7}_{-2}$ & $3^{+7}_{-2}$ & $124^{+29}_{-23}$ & $130^{+38}_{-26}$ & $1,000^{+550}_{-280}$ & $0.6^{+1.0}_{-0.3}$ & $39^{+9}_{-4}$ & $39^{+19}_{-20}$ & $-349 \pm 11$ \\
M22-L  & 2T$^{\dagger}$ & $4.1^{+1.0}_{-0.8}$ & $2^{+3}_{-2}$ & $95^{+17}_{-12}$ & $681^{+47}_{-46}$ & $\cdots$ & $\cdots$ & $34.7^{+1.6}_{-1.5}$ & $44^{+18}_{-20}$ & $-347 \pm 13$ \\
        & 1T & $2.05^{+0.19}_{-0.2}$ & $1.0^{+1.1}_{-0.9}$ & $197^{+11}_{-10}$ & $\cdots$ & $\cdots$ & $\cdots$ &$43^{+2}_{-2}$ & $52^{+19}_{-30}$ & $-319 \pm 17$ \\ 
\noalign{\vskip 0.6ex}
\tableline
\noalign{\vskip 0.6ex} 
       & 3T & $6^{+6}_{-2}$ & $2.5^{+5}_{-1.4}$ & $137^{+54}_{-31}$& $123^{+22}_{-22}$ & $773^{+360}_{-190}$ & $0.5^{+0.6}_{-0.3}$ & $49^{+10}_{-6}$ & $74^{+13}_{-15}$ & $-519 \pm 28$ \\
J21-L & 2T & $2.8^{+1.1}_{-1.0}$ &$1.5^{+2}_{-0.9}$ & $113^{+53}_{-22}$ & $711^{+110}_{-150}$ & $\cdots$ & $\cdots$ &$40^{+3}_{-2}$ & $85^{+12}_{-15}$ & $-505 \pm 28$ \\
       & 1T$^{\dagger}$ & $1.44^{+0.10}_{-0.08}$ &$0.6^{+0.7}_{-0.6}$ & $220^{+10}_{-8}$& $\cdots$ & $\cdots$ & $\cdots$ &$48^{+3}_{-3}$ & $85^{+13}_{-17}$ & $-506 \pm 28$ \\
\noalign{\vskip 0.6ex}
\tableline
\multicolumn{11}{c}{\text{Trailing Hemisphere}} \\
\tableline
\noalign{\vskip 0.6ex}
       & 3T & $0.75^{+0.2}_{-0.14}$ & $0.4^{+0.5}_{-0.3}$ & $193^{+943}_{-80}$ & $165^{+51}_{-52}$ & $667^{+390}_{-340}$ & $0.26^{+0.18}_{-0.2}$ & $38^{+7}_{-4}$ & $55^{+14}_{-18}$ & $-527 \pm 12$ \\
M22-T & 2T & $0.72^{+0.19}_{-0.12}$ & $0.4^{+0.5}_{-0.4}$ & $170^{+35}_{-44}$ & $540^{+270}_{-270}$ & $\cdots$ & $\cdots$ & $36^{+4}_{-2}$ & $56^{+15}_{-20}$ & $-528 \pm 12$ \\
       & 1T$^{\dagger}$ & $0.66^{+0.04}_{-0.03}$ &$0.3^{+0.4}_{-0.3}$& $213^{+6}_{-6}$ & $\cdots$ & $\cdots$ & $\cdots$ & $41.7^{+1.7}_{-1.6}$ & $57^{+15}_{-20}$ & $-529 \pm 13$ \\
\noalign{\vskip 0.6ex} 
\tableline
\noalign{\vskip 0.6ex} 
       & 3T & $1.2^{+1.4}_{-0.4}$ & $0.5^{+1.1}_{-0.3}$ & $132^{+140}_{-38}$& $115^{+45}_{-25}$ & $340^{+140}_{-100}$ & $0.3^{+0.4}_{-0.2}$ & $50^{+16}_{-8}$ & $26^{+20}_{-18}$ & $-601 \pm 17$ \\
J21-T & 2T & $0.9^{+0.6}_{-0.3}$ & $0.4^{+0.7}_{-0.3}$ & $116^{+30}_{-26}$ & $320^{+100}_{-120}$ & $\cdots$ & $\cdots$ &$46^{+9}_{-6}$ & $28^{+20}_{-19}$ & $-602 \pm 17$ \\
       & 1T$^{\dagger}$ & $0.66^{+0.06}_{-0.06}$ &$0.3^{+0.3}_{-0.2}$ & $152^{+10}_{-11}$ & $\cdots$ & $\cdots$ & $\cdots$ & $52^{+10}_{-7}$ & $26^{+20}_{-18}$ & $-604 \pm 18$ \\
\enddata
\tablecomments{Model comparison and retrieved parameters for each dataset and temperature-profile model. Entries report best-fit values retrieved from our minimization, with 68$\%$ credible intervals for SO$_2$ column density, surface pressure ($P_{\rm surf}$), \added{node temperature ($T_x$),} transition pressure ($P_{\mathrm{transition}}$), fractional coverage, and wind dispersion, together with LOOIC and its standard error (SE). \added{For multi-node models, the upper-node temperature is weakly constrained (limited sensitivity at $P\lesssim10^{-2}$~nbar) and should be interpreted as an effective upper-atmosphere parameter.} $\dagger$ marks in the ``Model'' column indicate the \added{lowest complexity model statistically indistinguishable from the minimum-LOOIC model in that dataset---those whose $\Delta$LOOIC relative to the minimum-LOOIC model is $\le$ the quadrature-sum SE. Entries with ``$\cdots$'' indicate that the parameter is not applicable for that model.}}

\end{deluxetable*}

\section{Discussion}
\label{sec:Discussion}

\begin{figure*}
    \centering
    \includegraphics[width=1\textwidth]{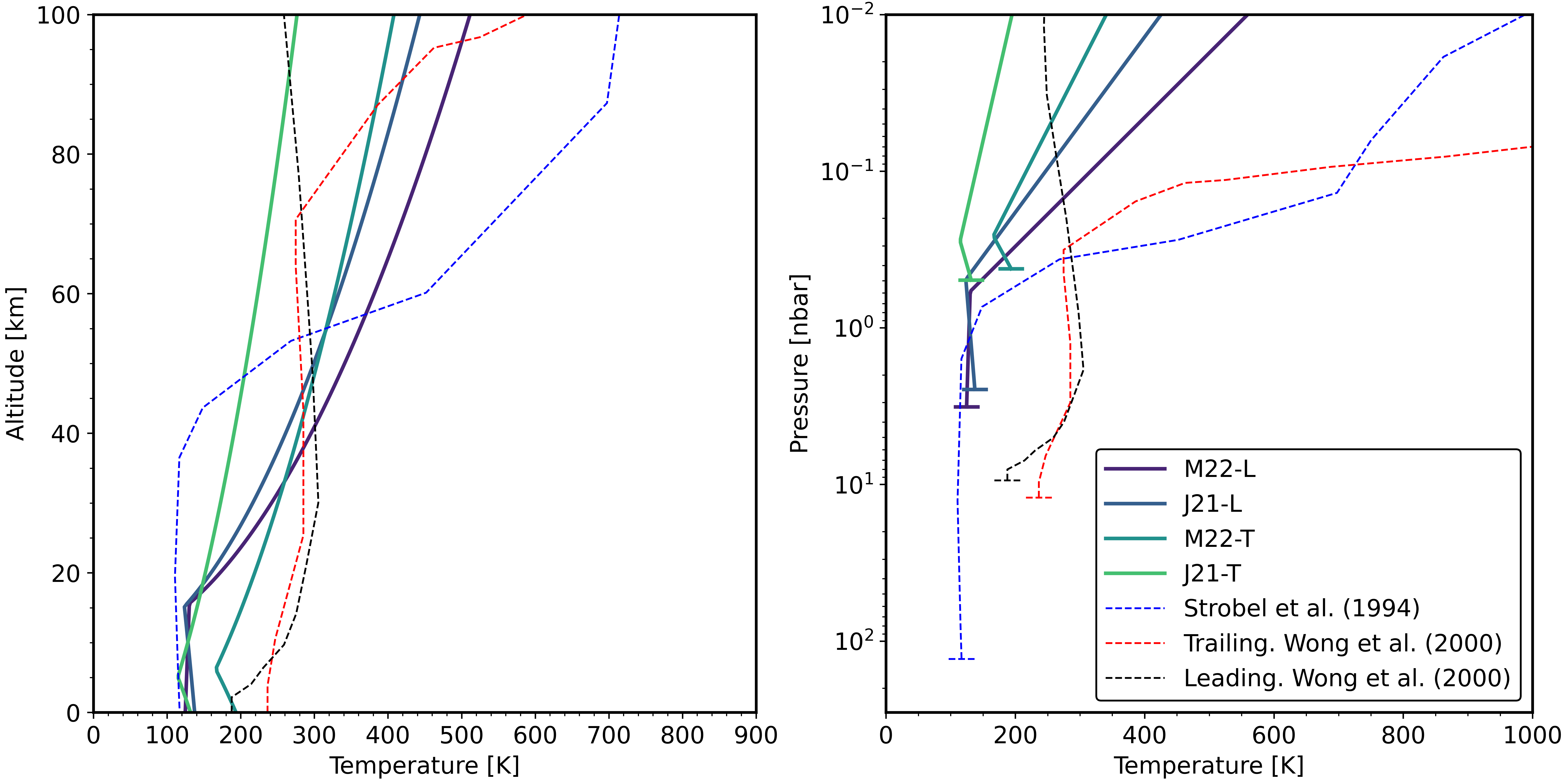}
    \caption{\added{Retrieved temperature-altitude profiles for Io from our four ALMA datasets (M22-L, J21-L, M22-T, and J21-T) each fit with the same three-node temperature parameterization (solid lines). The left panel shows the retrieved temperature profiles as a function of altitude (km), while the right panel shows the same profiles as a function of pressure (nbar). The horizontal bars in the right panel mark the surface pressure associated with each model. For context, dashed curves show representative model profiles from \citet{Strobel1994} and \citet{Wong2000}. Specifically for \citet{Strobel1994}, we plot the high-density dayside solutions from their model that includes solar, plasma, and Joule heating. For \citet{Wong2000}, we plot the high-density dayside solutions at western elongation (trailing hemisphere) and eastern elongation (leading hemisphere).}} 
    \label{fig:TP_model_comparison}
\end{figure*}

\subsection{Stability of Multi-Node Temperature Structures and Rationale for Adopting the 3T Model}\label{subsec: why3T}
Before we compare our retrieved temperature profiles with theoretical expectations, we emphasize which aspects of the retrieval are genuinely robust across datasets and parameterizations. Although the 1T, 2T, and 3T models can all produce acceptable fits---for the most part---in a purely statistical sense, they do not behave interchangeably. The clearest indication comes from the isothermal fits where, on both hemispheres, the retrieved 1T temperatures disagree at levels slightly or greatly exceeding their quoted uncertainties. Notably, the trailing side M22-T dataset yields an isotherm near 210~K, while the J21-T dataset yields roughly 155~K (despite the fact that the two datasets probe similar pressure ranges).

This inconsistency reflects the well-known degeneracy among temperature, SO$_2$ column density, and fractional coverage. With only a single temperature parameter available, many combinations of a cooler atmosphere with higher SO$_2$ column (and/or a larger fractional coverage), or a warmer atmosphere with lower SO$_2$ column (and/or a smaller fractional coverage), generate nearly identical line depths and relative line ratios. Therefore, the 1T retrieval likely follows the degeneracy direction rather than a physically meaningful temperature scale, allowing two satisfactory fits to report temperatures that differ by tens of kelvin.

We tested this behavior explicitly by re-fitting the M22-T spectra while fixing the isotherm to the value retrieved for J21-T. The spectral fit degraded only modestly, while the inferred column and coverage shifted in precisely the manner expected from the degeneracy. This behavior demonstrates that the mismatch between the two 1T temperatures is likely not atmospheric in origin but a modeling artifact. Consequently, the formal 1T uncertainties shown in Figures~\ref{fig: leading-temp-profiles} and~\ref{fig: trailing-temp-profiles} describe only the conditional width of the posterior once isothermality is imposed---they do not capture the broader physical uncertainty \added{arising from the strong correlations between temperature, SO$_2$ column density, and fractional coverage.} Once vertical structure is permitted, this disagreement across epochs disappears. Both the 2T and 3T parameterizations produce internally consistent temperature profiles within each hemisphere. Therefore, introducing even modest vertical flexibility removes the artificial spread produced by the 1T model and yields temperature structures that are stable across independent datasets.

The statistical behavior mirrors this pattern but with important nuance. As summarized in Table~\ref{tab:summary_results}, only the M22-L dataset exhibits statistical evidence against an isothermal atmosphere. Both 2T and 3T models improve the LOOIC by roughly thirty points relative to 1T. The 3T model yields the minimum LOOIC, though its difference from 2T is smaller than the combined standard error. The J21-L dataset shows the same ranking, but the differences fall below the quadrature-sum error and thus provide only weak discrimination. On the trailing hemisphere, all three temperature profile parameterizations are statistically indistinguishable in LOOIC. In other words, the multi-node models are favored where the data have vertical leverage and are neither required nor penalized strongly where the atmosphere is too thin to support strong discrimination.

These results motivate the use of multi-node temperature parameterizations as the most physically interpretable models. Allowing vertical structure removes the artificial spread produced by the 1T model and yields temperature profiles that are consistent across independent epochs. Additionally, the multi-node solutions can naturally reduce to simpler profiles when additional complexity is not required. Among these, the 3T parameterization provides the minimal flexibility needed to represent the structure anticipated by the theoretical models introduced in Section~\ref{subsec: theory-models}: a cool, quasi-isothermal lower layer followed by a thermospheric warming at higher altitudes, and---in some instances---a temperature minimum at the transition pressure (i.e. a mesopause). The 2T model cannot reproduce this mesopause-like structure. \added{In all retrievals, the best-fit 3T solution exhibits this mesopause-like curvature, aligning with theoretical expectations and further motivating its use for interpretation. We find no evidence that this structure is a fitting artifact because it appears independently in every dataset and persists under alternative temperature profile parameterizations explored in preliminary analyses. Moreover, the 1$\sigma$ temperature profile uncertainties do not always preserve the curvature, which suggests that the feature is data-supported but of finite significance rather than structurally imposed by the model.}

For these reasons, we adopt the 3T model as the fiducial framework for interpretation and comparison with theoretical temperature profile predictions. In the sub-sections that follow, we focus primarily on the leading hemisphere because its higher SO$_2$ columns and stronger rotational transitions distribute their contribution function peaks across a broader pressures range---providing the vertical leverage needed to distinguish among temperature profile parameterizations. By contrast, the trailing hemisphere is fundamentally sensitivity-limited because its low SO$_2$ column pushes optical-depth unity to low altitudes and compresses the contribution functions into a narrow pressure band, rendering the various temperature parameterizations statistically indistinguishable.

\subsection{Io's Leading Hemisphere Vertical Temperature Profile}\label{subsec: Io-TP}
Our leading hemisphere retrievals under the 3T parameterization reveal temperature structures that are compatible with the mesopause-like minimum predicted by \citet{lellouch1990io, Lellouch1992} and quantified by \citet{Strobel1994}. In the J21-L dataset, a shallow temperature minimum at the transition pressure is retrieved, while in M22-L a similar feature is allowed by the uncertainties but not strongly required by the data. In both datasets, the lower atmosphere remained quasi-isothermal up to about $\sim\!0.5$~nbar ($\sim$15~km in altitude), with temperatures tightly constrained to $\sim\!124$~K for M22-L and $\sim\!137$~K for J21-L. Above this level, the temperature rose by several hundred kelvin, reaching values of $\sim\!$400-600~K by 10$^{-2}$~nbar ($\sim$100~km in altitude). This retrieved structure maps closely onto the canonical framework laid out by \citet{Strobel1994}: at surface pressures above $\sim\!10$~nbar, collisional radiative cooling by SO$_2$ rotational and non-LTE vibrational bands depresses the temperature of the lower atmosphere, producing a cold point near the nanobar level, while Joule heating due to the corotating plasma bound to the Jovian magnetosphere dominates at subnanobar pressures and drives the temperature upward. The commonalities between our retrieved profiles and \citet{Strobel1994}'s model are best visualized in Figure~\ref{fig:TP_model_comparison}, where we compare \citet{Strobel1994} with our 3T retrievals for both hemispheres under a uniform parameterization. For context, we also include the dayside model profiles of \citet{Wong2000}, discussed later in Subsection~\ref{subsec: hemi}.

The most notable difference between our leading hemisphere retrieved TPs and those from \citet{Strobel1994} is that our retrieved leading hemisphere surface pressures of $\sim$3~nbar are somewhat lower than the $\sim$10~nbar \added{surface pressure} threshold cited by \citet{Strobel1994} for mesopause formation, but the correspondence in vertical structure is clear. In \citet{Strobel1994}'s analysis, a \added{$\sim$100~nbar modeled} surface pressure yields a quasi-isothermal temperature extending up to $\sim$1~nbar. Naturally, since the non-LTE vibrational SO$_2$ cooling that likely forms this quasi-isothermal structure scales with $P^2$ \citep{Lellouch1992}, lower \added{surface} pressures reduce the effectiveness of this cooling and we'd expect heating from above to take over sooner---which is consistent with our retrievals. Over the sub-nbar range, the inferred thermospheric slope is consistent with \citet{Strobel1994} within uncertainties. 

The pressure (altitude) of the inflection/cold point and the thermospheric slope provide complementary diagnostics. A cold point at higher pressure (deeper) indicates weaker surface support and reduced collisional cooling, whereas a higher-altitude cold point implies a denser column in which SO$_2$ cooling remains effective over a larger vertical extent. The thermospheric slope mainly reflects ionospheric/Joule heating \citep{Strobel1994}. Inverting these together constrains the combination of dayside surface pressure (sublimation supply) and ionospheric conductance, offering a concrete benchmark for forward models.

By contrast to \citet{Strobel1994}'s model, 3-D rarefied gas dynamics models of Io's atmosphere that explicitly track sublimation and volcanic sources (e.g., \citealt{walker2010comprehensive, Gratiy2010}) do not produce such a quasi-isothermal lower layer. Instead, their translational and rotational temperatures increase monotonically with altitude, while the $\nu_2$ vibrational mode rapidly departs from LTE and remains cooler than the kinetic temperature. The absence of a cold point in these DSMC results likely reflects their emphasis on rarefied flow physics, global transport, and plume-driven inhomogeneity, which may smooth vertical gradients that appear sharply in 1-D radiative-conductive models. 

Our mm/sub-mm leading hemisphere retrievals therefore provide a new observational benchmark: atmospheric models must simultaneously capture the strong SO$_2$ cooling in the lower atmosphere and the dominant Joule heating from above, while also accounting for non-LTE processes and large-scale circulation. While this agreement with the canonical radiative-conductive framework is compelling, several caveats must be considered. First, at the lowest pressures probed, our LTE assumption means that retrieved temperatures should be treated as effective excitation temperatures rather than strict kinetic values. However, we emphasize that for pressures greater than $P\sim0.02$~nbar (reported by \citet{Gratiy2010} as about 50~km in their adopted profile), the rotational and kinetic temperatures follow closely. Second, the mapping between line opacity and pressure rests on the assumption of SO$_2$ dominance and hydrostatic balance, which may not hold uniformly across Io's heterogeneous atmosphere. Third, our spectra are extracted at disk-center and parameterized with a fractional coverage term; this captures unresolved inhomogeneity but does not resolve longitudinal or latitudinal structure. These limitations do not remove the need for vertical structure in Io's temperature profile, but they do introduce uncertainties in the precise pressure/altitude and magnitude of the retrieved inflection/cold point.

\subsection{Cross-Wavelength Context and Implications}
Although our retrievals are based on mm/sub-mm rotational lines, the vertical flexibility of the 3T parameterization is relevant beyond this wavelength regime. As discussed in Section~\ref{subsec: Temp-limitations} and reviewed in \citet{Lellouch2015} and \citet{dePater2023}, temperature estimates obtained across observational wavelengths have historically differed, in part because each diagnostic samples different pressure ranges and relies on distinct assumptions regarding winds and non-LTE processes. \added{With the additional vertical flexibility provided by the 3T parameterization, the lower atmosphere temperatures ($\sim$124-137~K for most 3T fits) are consistent with interpretations of the 19$\mu$~m SO$_2$ absorption bands, which required atmospheric temperatures below 170~K \citep{Giles2024} and were best fit with temperatures of $108\pm18$~K \citep{Tsang2012}.} At high altitudes, recent JWST detections of sulfur emission (1.08 $\&$ 1.13 micrometers) require atmospheric temperatures of order 1,700 K above $\sim\!$200-300~km \citep{dePater2025first}. These high-altitude temperature values are consistent with earlier heating predictions discussed in Section~\ref{subsec: theory-models}, but they exceed the temperature values retrieved at lower altitudes by infrared and mm/sub-mm observations. \added{The 3T models retrieved here naturally support this qualitative vertical picture because they allow temperatures to increase with altitude while remaining cold in the lower atmosphere (accommodating both the cool lower atmosphere constraints and the hotter upper atmospheric temperatures inferred from JWST).} A full reconciliation across wavelengths is outside the scope of this work, but the 3T parameterizations retrieved here offer testable vertical templates for future multi-wavelength analyses to evaluate whether the curvature we infer helps contextualize these cross-wavelength temperature measurements.

In addition to the vertical structure retrieved under the 3T parameterization, the isothermal (1T) fits themselves provide useful cross-wavelength context because they represent the effective LOS-averaged temperatures sampled by the rotational lines. The 1T temperatures retrieved on the leading hemisphere (195-220~K) and on the trailing hemisphere (150-210~K) fall within the 150-220~K range inferred from 4~$\mu$m spectroscopy and near the $\sim\!170\pm20$~K mean value reported by \citet{Lellouch2015}. However, as discussed in Section~\ref{subsec: why3T}, these 1T temperatures vary between epochs and should only be viewed as degenerate LOS-averaged solutions rather than unique physical estimates of the atmosphere. Their agreement with the 4~$\mu$m results is encouraging at the level of broad consistency, but it does not resolve the intrinsic degeneracies that allow different datasets to yield discrepant 1T temperatures (despite the similar pressure coverage). This reinforces why our multi-node profiles are required for any physically interpretable comparison across wavelengths.

\subsection{Trailing Hemisphere Limitations}\label{subsec: trailing-discussion}
Our trailing hemisphere retrievals provide a useful control because they represent the limiting case in which the data contain minimal vertical temperature leverage. The low SO$_2$ column and accompanying lower signal-to-noise force $\tau_\nu\!\sim\!1$ toward higher pressures, compressing the line-forming region downward such that all the targeted transitions weight a comparatively narrow atmospheric layer rather than the broader 0-30~km span accessed on the leading side. With this limited vertical range on the trailing side, additional temperature nodes receive minimal constraints from higher, potentially warmer layers, and the 1T, 2T, and 3T models therefore perform equivalently. Such nodes on the trailing hemisphere neither requires nor rules out an inflection or cold point---they simply do not probe the pressures where such structure would matter radiatively. In this limited-leverage regime, quantitative temperatures in the more flexible parameterizations are likely not physically meaningful. Therefore, temperature inconsistencies, such as the elevated lower node temperature seen in the 3T M22-T retrieval in Figure~\ref{fig: trailing-temp-profiles}, are best understood as an artifact of over-parameterization---consistent with the LOOIC results showing that the spectra do not justify additional vertical structure on the trailing hemisphere.

Although the node-level temperatures are not individually meaningful under these conditions, we do note that the 3T temperature profiles still exhibit some quasi-isothermal structure with a mild upward turn, as shown in Figure~\ref{fig: trailing-temp-profiles}, but the magnitude and altitude of that rise are weakly constrained. Very strong upper nodes are disfavored, as these would overproduce line-core flux in some transitions, but moderate thermospheric heating remains permitted within the uncertainties. Additionally, since SO$_2$ non-LTE vibrational cooling scales strongly with density ($\propto P^2$), we expect the trailing hemisphere---with its systematically lower column densities---to be less efficient at sustaining a pronounced cold point. This may explain why the leading hemisphere, with larger SO$_2$ columns, point toward curvature in the temperature profile, while the trailing hemisphere spectra remain consistent with isothermality.   

Overcoming this sensitivity limitation will require observations that include transitions that recover vertical leverage at lower pressures. As such, ALMA observations covering lines with a wider span in excitation energies could be particularly valuable, since they would stagger the contribution functions and help de-correlate column density from temperature structure. In our case, the Band 7 trailing-hemisphere dataset already demonstrates this: by targeting a larger set of lines with differing $E_{\rm low}$, the contribution functions are more vertically separated than in the Band 8 dataset, which relied on a tighter spread of excitation energies (see Figure~\ref{fig: trailing-contrib-profiles} in Appendix~\ref{sec:AppendixContrib}). As a result, the Band 7 spectra provided better vertical leverage and begin to reveal weak curvature and decouple column density from temperature structure. Future campaigns that combine additional ALMA bands or multi-epoch stacking will be essential for determining whether the trailing hemisphere's vertical structure is similar to that of the leading hemisphere.

\subsection{Hemispheric Differences Between Temperature Profiles}\label{subsec: hemi}
As mentioned earlier, Figure~\ref{fig:TP_model_comparison} shows our retrieved 3T temperature profiles alongside modeled profiles from \citet{Strobel1994} and \citet{Wong2000}. We only plot up to the altitude\added{/pressure} which our spectra are sensitive to (up to $\sim$100~km\added{/10$^{-2}$~nbar}). On the \emph{leading} hemisphere, both datasets show a well-defined quasi-isothermal region up to $\sim$15~km followed by a steep thermospheric rise. As discussed in Subsection~\ref{subsec: Io-TP}, this structure matches the classic radiative-conductive model of \citet{Strobel1994} (mesopause capped by plasma/Joule heating above; black dashed line in~Figure~\ref{fig:TP_model_comparison}). In contrast, the \citet{Wong2000} high-density solutions predict only quasi-isothermal atmosphere---with no sharp thermospheric rise---on the leading dayside at eastern elongation (black dashed line in~Figure~\ref{fig:TP_model_comparison}), while the sharp turnover appears only in their trailing dayside western elongation solution (red dashed line in~Figure~\ref{fig:TP_model_comparison}). Our leading profiles therefore exceed Wong's leading side expectations and resemble, in shape, the turnover they associate with the trailing hemisphere (a quasi-isothermal, cold lower atmosphere followed by a rise in temperature at higher altitudes). This difference can be reconciled if ionospheric/Joule heating on the leading side is more effective than \citet{Wong2000} assumed.

On the \emph{trailing} hemisphere, the retrieved 3T profiles are qualitatively similar with the leading hemisphere retrievals, but they are flatter with a more muted upturn, as shown in~Figure~\ref{fig:TP_model_comparison}. As discussed in Subsection~\ref{subsec: trailing-discussion}, this owes primarily to sensitivity limits from the systematically lower SO$_2$ columns on the trailing side: the contribution functions cluster near the surface, making 1T/2T/3T models statistically indistinguishable and leaving the sub-nanobar regime largely unconstrained. The absence of a detected strong thermospheric rise on the trailing side is therefore best understood as a sensitivity limitation.

\section{Conclusion}
\label{sec:Conclusion}
We used multi-transition ALMA spectroscopy of SO$_2$ to retrieve Io's first vertically resolved millimeter/submillimeter temperature profiles. On the leading hemisphere, the data point toward vertical structure. Under the 3T temperature profile parameterization, we obtain a quasi-isothermal lower atmosphere at $\sim$124-137~K up to $\sim$0.5~nbar ($\sim$15~km in altitude), followed by a rise in temperature reaching $\sim\!$400--600~K by 10$^{-2}$~nbar ($\sim$100~km in altitude). This pattern is consistent with the vertical structure predicted by radiative-conductive models \citep{Lellouch1992, Strobel1994}, but we do not claim a uniquely resolved temperature minimum (mesopause). Moreover, since our spectra primarily weight pressures greater than $\sim$10$^{-2}$~nbar and rotational LTE \added{is not expected to hold at pressures below $P\sim0.02$~nbar (corresponding to $\sim$50~km altitude, depending on the assumed surface pressure; \citealt{Gratiy2010})}, the high-altitude temperature values should be read as loosely constrained effective rotational temperatures rather than direct measurements of the kinetic temperature. While a full reconciliation of historical temperature discrepancies across observational wavelengths is beyond the scope of this work, the multi-node profiles retrieved here provide testable vertical templates that can be used in future analyses to evaluate whether curvature in Io's temperature profile helps contextualize those earlier measurements.

In contrast to our leading hemisphere retrievals, the trailing hemisphere spectra remained statistically consistent with an isothermal atmosphere---a consequence of lower SO$_2$ columns compressing $\tau_\nu\!\sim\!1$ toward the surface and limiting vertical leverage. We note, although, that the more complex, multi-node profiles retrieved for the trailing hemisphere qualitatively matched those retrieved from the leading hemisphere. Across epochs, we recovered systematically higher SO$_2$ columns on the leading side (few $\times 10^{16}\ \mathrm{cm^{-2}}$) than the trailing side ($\sim10^{15}$-$10^{16}\ \mathrm{cm^{-2}}$), with fractional beam coverage of $\sim$35-50$\%$ and sub-beam velocity dispersions of $\sim$25-85~m~s$^{-1}$ for both hemispheres.

Methodologically, three components were essential to mitigate classic degeneracies between temperature, column density, and atmosphere dynamics---simultaneous fitting of lines spanning a range of lower-state energies; LOS Doppler map informed forward modeling plus an explicit sub-beam dispersion term to correctly interpret the line shape; and robust Bayesian inference with channel-level PSIS-LOO-CV to determine the level of model complexity that is statistically warranted. Together these components deliver the vertical sensitivity needed to identify vertical structure on the leading hemisphere and to diagnose why the trailing side is currently sensitivity-limited rather than intrinsically isothermal. Our results also provide a quantitative benchmark for the thermal energy balance of Io's atmosphere---and a template for coupling atmospheric layered structures to magnetospheric forcing on other volcanically or tidally active worlds.
 
Looking forward, our analysis points to these key steps: (i) joint ALMA Band campaigns to stagger contribution functions and extend low-pressure leverage; (ii) contemporaneous UV--near-IR measurements to constrain the thermospheric temperatures shaped by plasma and Joule heating; (iii) non-LTE radiative-transfer forward models (rotational + vibrational) embedded in the retrieval to better characterize the thermospheric slope; and (iv) targeted trailing-side observations at phases of elevated SO$_2$ support (e.g., near perihelion) to test for a vertical structure under high-density conditions. Sharper constraints on the thermospheric slope will pinpoint the altitude of energy deposition as well as the exobase temperature that governs scale heights and escape. 

\section{Data and Materials Availability}
\added{The ALMA observations analyzed in this study correspond to the project codes ADS/JAO.ALMA\#2019.1.00216.S and ADS/JAO.ALMA\#2021.1.00849.S, which are publicly accessible through the ALMA Science Archive at \url{https://almascience.nrao.edu/aq/?projectCode=2019.1.00216.S} and \url{https://almascience.nrao.edu/aq/?projectCode=2021.1.00849.S}, respectively. In addition, the corner plot figures, along with all data required to reproduce the main temperature profile figures, are available via a Zenodo repository (DOI: \href{https://doi.org/10.5281/zenodo.18342667}{10.5281/zenodo.18342667}). The data are provided in CSV format; for each retrieval, two files are included, one tabulated as a function of pressure and the other as a function of altitude.}

\begin{acknowledgments}
This paper makes use of the following ALMA data: ADS/JAO.ALMA\#2019.1.00216.S and ADS/JAO.ALMA\#2021.1.00849.S. ALMA is a partnership of ESO (representing its member states), NSF (USA) and NINS (Japan), together with NRC (Canada), NSTC and ASIAA (Taiwan), and KASI (Republic of Korea), in cooperation with the Republic of Chile. The Joint ALMA Observatory is operated by ESO, AUI/NRAO and NAOJ. The National Radio Astronomy Observatory and Green Bank Observatory are facilities of the U.S. National Science Foundation operated under cooperative agreement by Associated Universities, Inc.

KdK acknowledges support from the National Science Foundation (NSF) under Grant No. 2238344 through the Faculty Early Career Development Program. IdP and SLC acknowledge funding from NASA SSW Grant 80NSSC24K0306, as a sub-awardee of the lead campus, University of Texas at Austin. TNP acknowledges support from the Caltech Rosenbaum-Faber Family Graduate Fellowship. We thank Zachariah Milby (\href{https://orcid.org/0000-0001-5683-0095}{ORCiD: 0000-0001-5683-0095}) for his expertise in figure design, layout, and data visualization. We thank the anonymous reviewers for their valuable feedback and constructive comments, which helped improve the clarity and quality of this manuscript.
\end{acknowledgments}

\begin{appendix}
\section{Impact of Assumed Surface Temperature on Atmospheric Retrieval}
\label{sec:AppendixSurfaceTemp}
To assess the sensitivity of our retrieval framework to the assumed surface temperature, we conducted a targeted comparison using the M22-L dataset. We performed three retrievals, each using a fixed surface temperature of 100, 115, or 130~K, while keeping all other modeling assumptions identical. In all cases, we adopted a surface emissivity of 0.8, consistent with previous millimeter studies of Io \citep{deKleer2024, dePater2020b}. Additionally, for all retrievals, we used the three-node (3T) temperature profile because this model had the most flexibility and best captured the vertical structure. This consistency allowed any differences to be attributed solely to the fixed surface temperature. 

\begin{figure*}
    \centering
    \includegraphics[width=\textwidth]{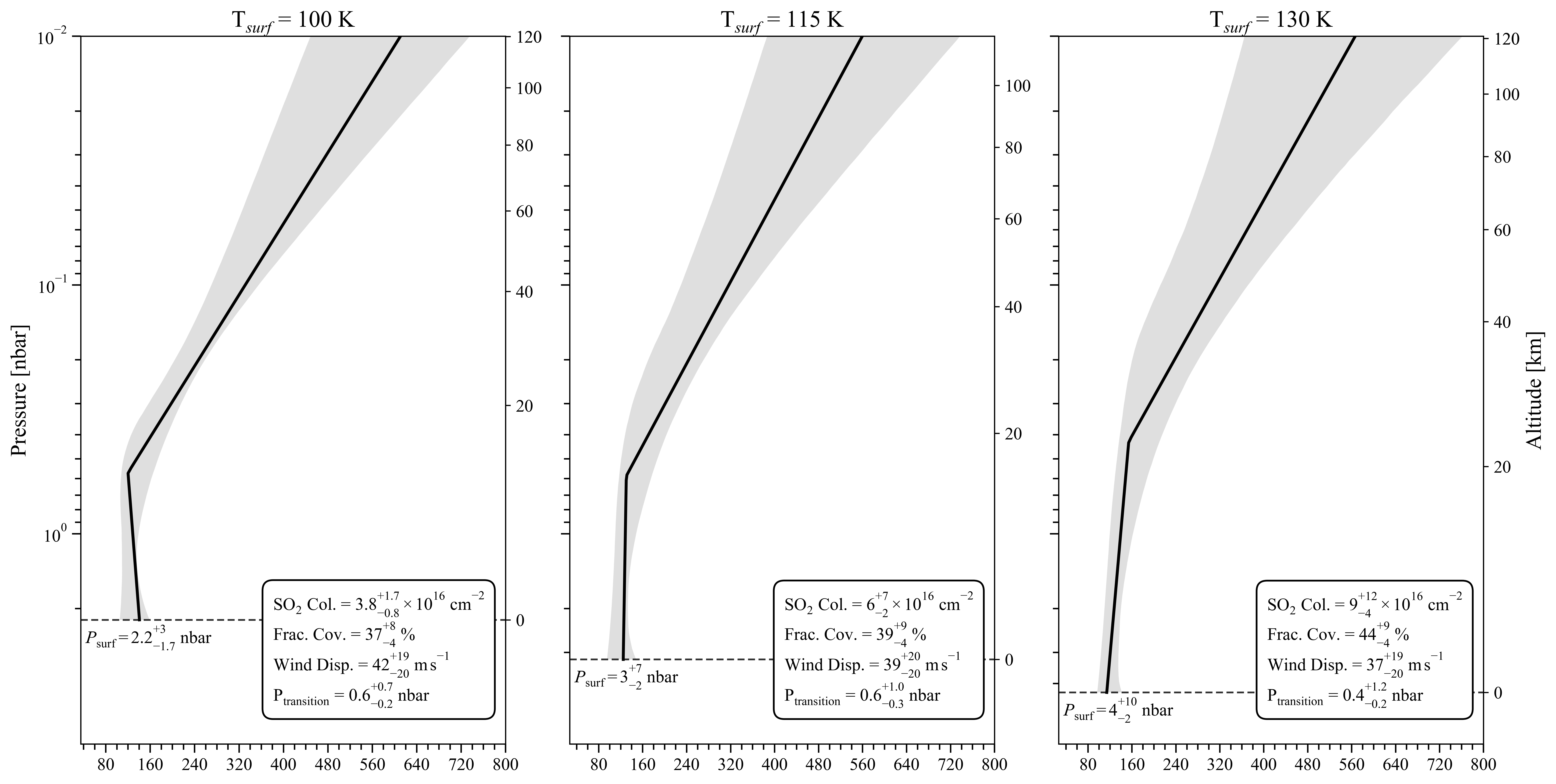}
    \caption{\textbf{Effect of Fixed Surface Temperature on Retrieved Atmospheric Structure.} The three columns correspond to retrievals from the M22-L dataset performed with fixed surface temperatures of 100~K (left), 115~K (center), and 130~K (right), each assuming an emissivity of 0.8. The panels show the retrieved 3T temperature profiles as a function of pressure. The black curve is the best-fit TP; the gray band marks the 1$\sigma$ credible interval. The legend for each profile provides the best-fit SO$_2$ column density, fractional coverage, wind dispersion, and transition pressure (with $1\sigma$ errors). Left axes give pressure; right axes show the corresponding altitude. The dashed horizontal black line indicates the modeled surface pressure, quoted with uncertainties calculated from the upper and lower bounds of the SO$_2$ column density.}
    \label{fig:Tsurf-composite}
\end{figure*}

All three models produced nearly indistinguishable fits to the observed SO$_2$ lines, with LOOIC scores that were statistically equivalent: LOOIC = $-349 \pm 12$ for the T$_{surf}$ = 100~K model, LOOIC = $-349 \pm 11$ for the T$_{surf}$ = 115~K model, and LOOIC = $-350 \pm 11$ for the T$_{surf}$ = 130~K model. Additionally, as shown in Figure~\ref{fig:Tsurf-composite}, the retrieved temperature profiles, SO$_2$ column densities, fractional coverage, wind dispersion, and transition pressure varied between retrievals; however, these differences remained within their respective uncertainties. The overall temperature structure was roughly consistent across all cases: a quasi-isothermal atmosphere up to about $\sim\!0.5$~nbar, after which the temperature increased to $\sim\!600$~K. These retrievals highlight that while the choice of surface temperature can lead to small shifts in individual parameters, it does not substantially alter the physical interpretation of the atmosphere's vertical structure. That is, there are always going to be an ensemble of profiles that fit the data, and these all fall within the errors of the other profiles. We therefore conclude that fixing the surface temperature to 115~K introduced no significant bias and is an appropriate choice for our retrievals. 

\section{Contribution Function Analysis}
\label{sec:AppendixContrib}
\subsection{Leading Hemisphere}
\begin{figure*}
    \centering
    \includegraphics[width=\textwidth]{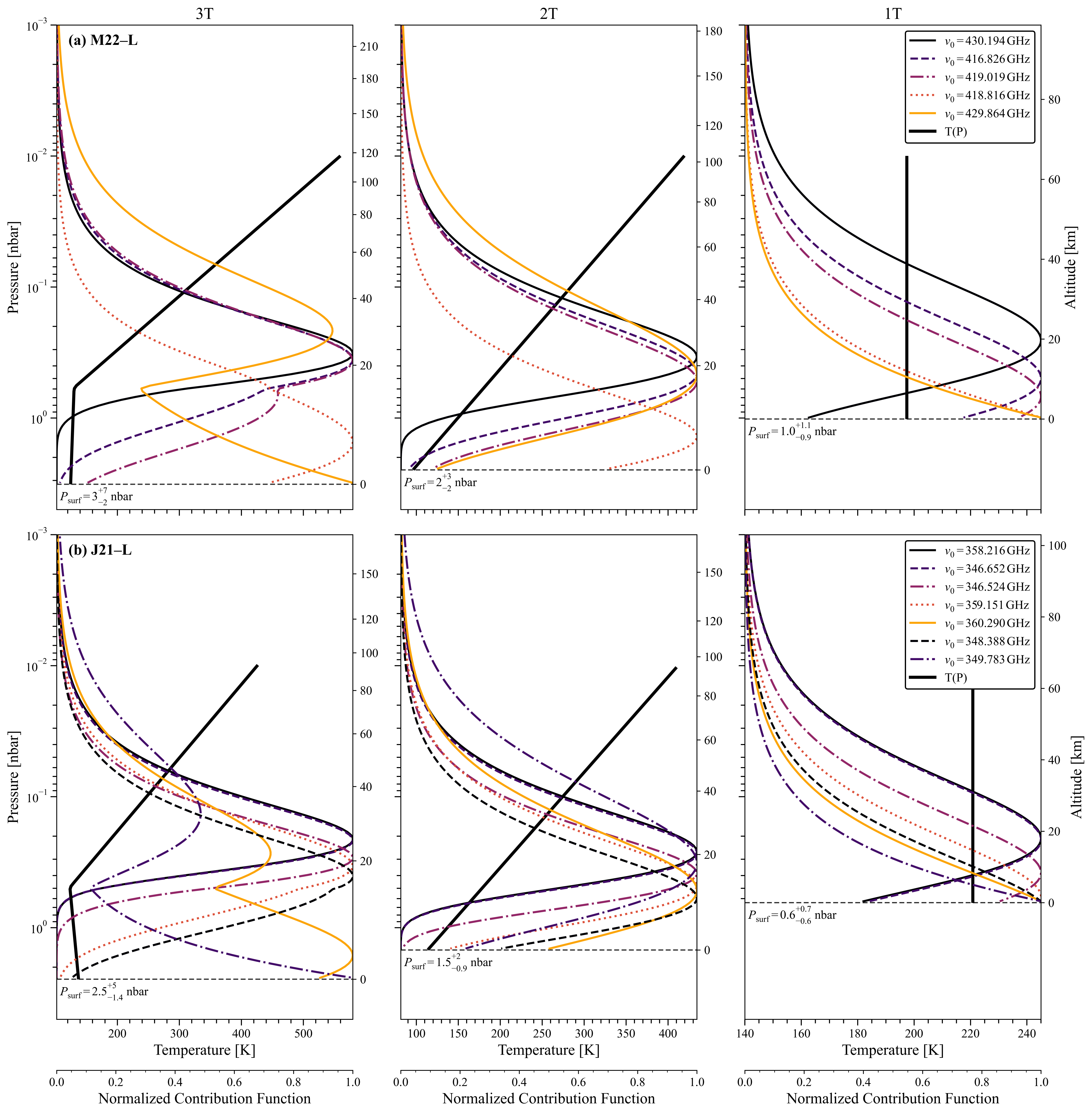}
    \caption{\textbf{Contribution functions and retrieved temperature profiles for leading hemisphere datasets: M22-L (a, top row) and J21-L (b, bottom row).} Columns show the 3-node (3T), 2-node (2T), and 1-node (1T) parameterizations. Colored curves are the normalized line center, $\nu_0$, contribution functions for the SO$_2$ transitions listed in the legend (frequencies in GHz); larger contribution function values indicate layers that contribute most to the emergent flux. The thicker, solid line is the best fit TP. Left axes give pressure; right axes show the corresponding altitude. The dashed horizontal black line indicates the modeled surface pressure, quoted with uncertainties calculated from the upper and lower bounds of the SO$_2$ column density.}
    \label{fig: leading-contrib-profiles}
\end{figure*}

Figure~\ref{fig: leading-contrib-profiles} shows how the contribution functions change with the corresponding temperature profile for both leading hemisphere datasets M22-L and J21-L. Because each function scaled as $B_{\nu}(T(p))e^{\frac{-\tau_{\nu}}{\mu_0}}\frac{d\tau_{\nu}}{d\ln(p)}$ where $\mu_0$ is the cosine of the emission angle between the LOS and the surface normal, altering $T(p)$ modified the source term, the temperature-dependent opacity, and the transmissivity through overlying layers. We analyze both datasets at once given their similarities. In the 1T, isothermal case, the contribution functions are strongly overlapped near the surface, with several truncated at the lower boundary by the relatively low retrieved surface pressure. With the 2T profile, a single monotonic slope shifts all contribution functions upward, but the peaks remain largely co-located. In this case, the vertical leverage is improved but still limited. The 3T profile introduces layered structure---quasi-isothermal through the first few scale heights then steeply rising afterwards---which preserves strong high-pressure weighting for the deeper-forming lines while enhancing the low-pressure sensitivity of the others. The set becomes clearly staggered from the surface to $\sim\!30$~km. This behavior is most evident in the weak, high-$E_{\rm low}$ transitions at 429.864 GHz in M22-L (Figure~\ref{fig: leading-contrib-profiles}, row a; solid yellow) and at 360.290 and 349.783 GHz in J21-L (Figure~\ref{fig: leading-contrib-profiles}, row b; solid yellow and dashed purple, respectively). These transitions still peak near the surface because they only reach appreciable optical depth in the denser layers. At lower pressures, above the inversion, these transitions develop a secondary, smaller maximum. By contrast, the lower-$E_{\rm low}$ and intrinsically stronger transitions reach optical depths of order unity at much lower pressures (e.g., transitions at 430.194 GHz in M22-L and at 358.216 GHz in J21-L; solid black line in Figure~\ref{fig: leading-contrib-profiles} for both rows). 

These patterns qualitatively explain why the 3T parameterization yields the lowest nominal LOOIC for the leading hemisphere, even though its advantage over 2T is not statistically distinguishable once uncertainties are included. What limits the 1T and 2T temperature profiles is their inability to satisfy transitions whose contribution functions diverge vertically. With the 1T profile, the temperature is the same at all pressures, so any attempt to help the fit to the lines that sense lower pressures simultaneously spoiled the fit to the deeper forming lines---the fit compensated by trading off temperature, column, fractional coverage, and/or wind dispersion, but cannot satisfy all the transitions once their vertical sensitivities differ. With the 2T profile, the fit was forced into the same compromises, just slightly relaxed given the additional temperature node. Allowing the layered structure in the 3T profile did the best of constraining these parameter degeneracies. Relative to the upper atmosphere, the cool base anchored the depths and core widths of the deep-forming lines, fixing the lower-atmosphere temperature and the column they require. The hot upper node added the extra flux in the low-pressure layers where the other lines retain sensitivity, so the fit no longer needed to inflate other atmospheric parameters (such as the wind-dispersion term) to accommodate accordingly.

\subsection{Trailing Hemisphere}
\begin{figure*}
    \centering
    \includegraphics[width=\textwidth]{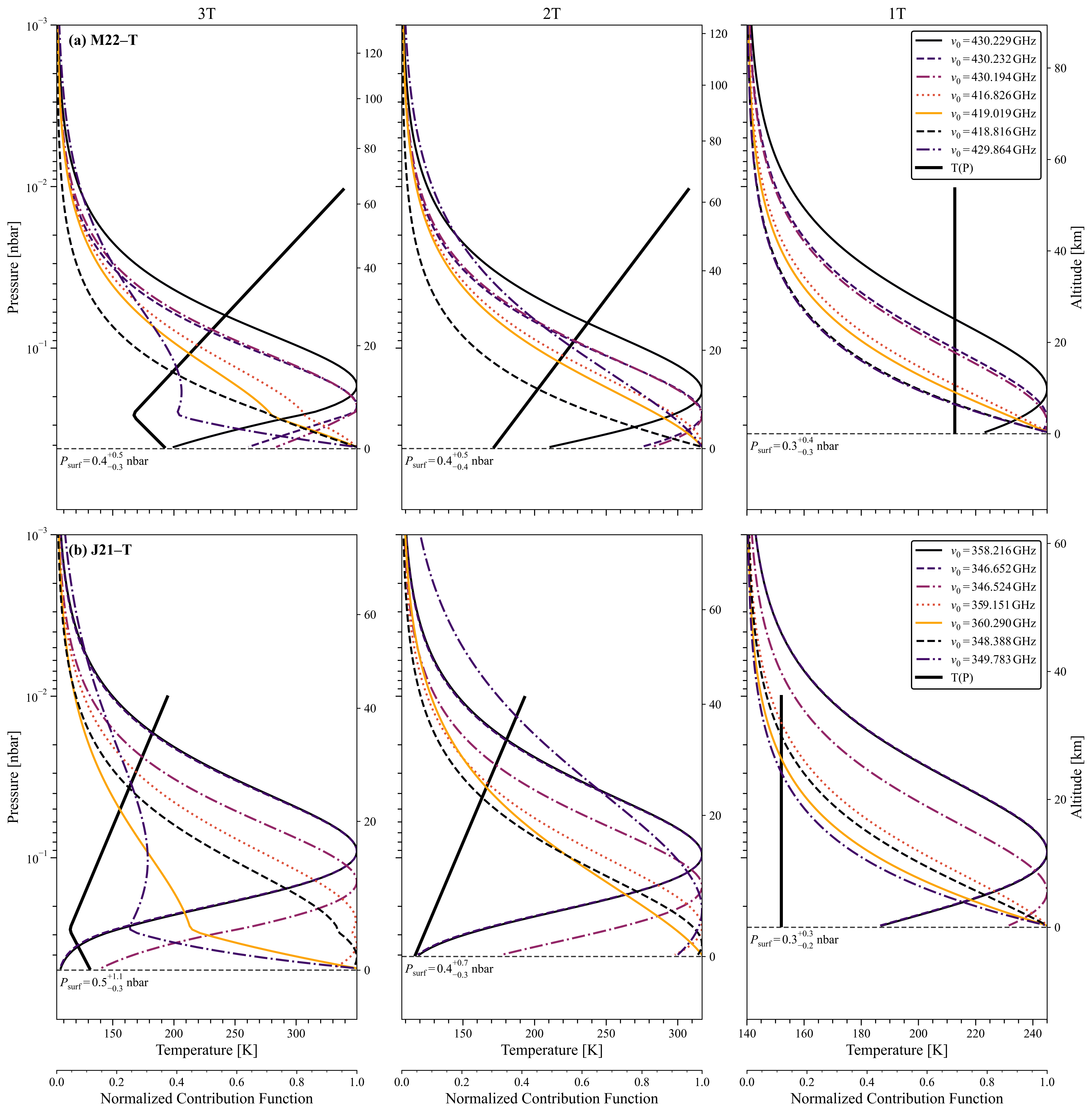}
    \caption{\textbf{Contribution functions and retrieved temperature profiles for trailing hemisphere datasets: M22-T (a, top row) and J21-T (b, bottom row).} Columns show the 3-node (3T), 2-node (2T), and 1-node (1T) parameterizations. Colored curves are the normalized line center, $\nu_0$, contribution functions for the SO$_2$ transitions listed in the legend (frequencies in GHz); larger contribution function values indicate layers that contribute most to the emergent flux. The thicker, solid line is the best fit TP. Left axes give pressure; right axes show the corresponding altitude. The dashed horizontal black line indicates the modeled surface pressure, quoted with uncertainties calculated from the upper and lower bounds of the SO$_2$ column density.}
    \label{fig: trailing-contrib-profiles}
\end{figure*}

Figure~\ref{fig: trailing-contrib-profiles} shows how the contribution functions change with the corresponding temperature profile for both trailing hemisphere datasets M22-T and J21-T. Across TP parameterizations, the contribution functions remain tightly confined near the surface and moving to more complex models only nudges some peaks slightly upward without creating distinct high-altitude weight (unlike the one in the leading hemisphere). The weaker lines and lower signal-to-noise that accompanied the low column on the trailing side further reduced vertical discrimination, so the retrievals produced mutually consistent temperature profiles (except for the 1T model, as discussed in Section~\ref{subsec: why3T}), with retrieved parameters effectively unchanged across temperature parameterizations. For M22-T, across temperature parameterizations, the SO$_2$ column varied only from $\sim\!0.66$ to $0.75\times10^{16}~\mathrm{cm^{-2}}$, fractional coverage from 36$\%$ to 42$\%$, and wind dispersion remained around 56~$\mathrm{m~s^{-1}}$---all differences well within the 1$\sigma$ intervals. J21-T showed a similar pattern, with somewhat larger column variation across parameterizations but indistinguishable LOOIC. Notably, the J21-T dataset provided slightly better vertical leverage than M22-T, since it included a wider spread of SO$_2$ transitions with differing excitation energies; this produced a modest separation in the contribution functions and hints at weak curvature, though the effect remains too small to yield a statistically distinguishable model preference.

\section{Additional Figures}
\label{sec:AppendixFigures}
\begin{figure*}
    \centering
    \includegraphics[width=\textwidth]{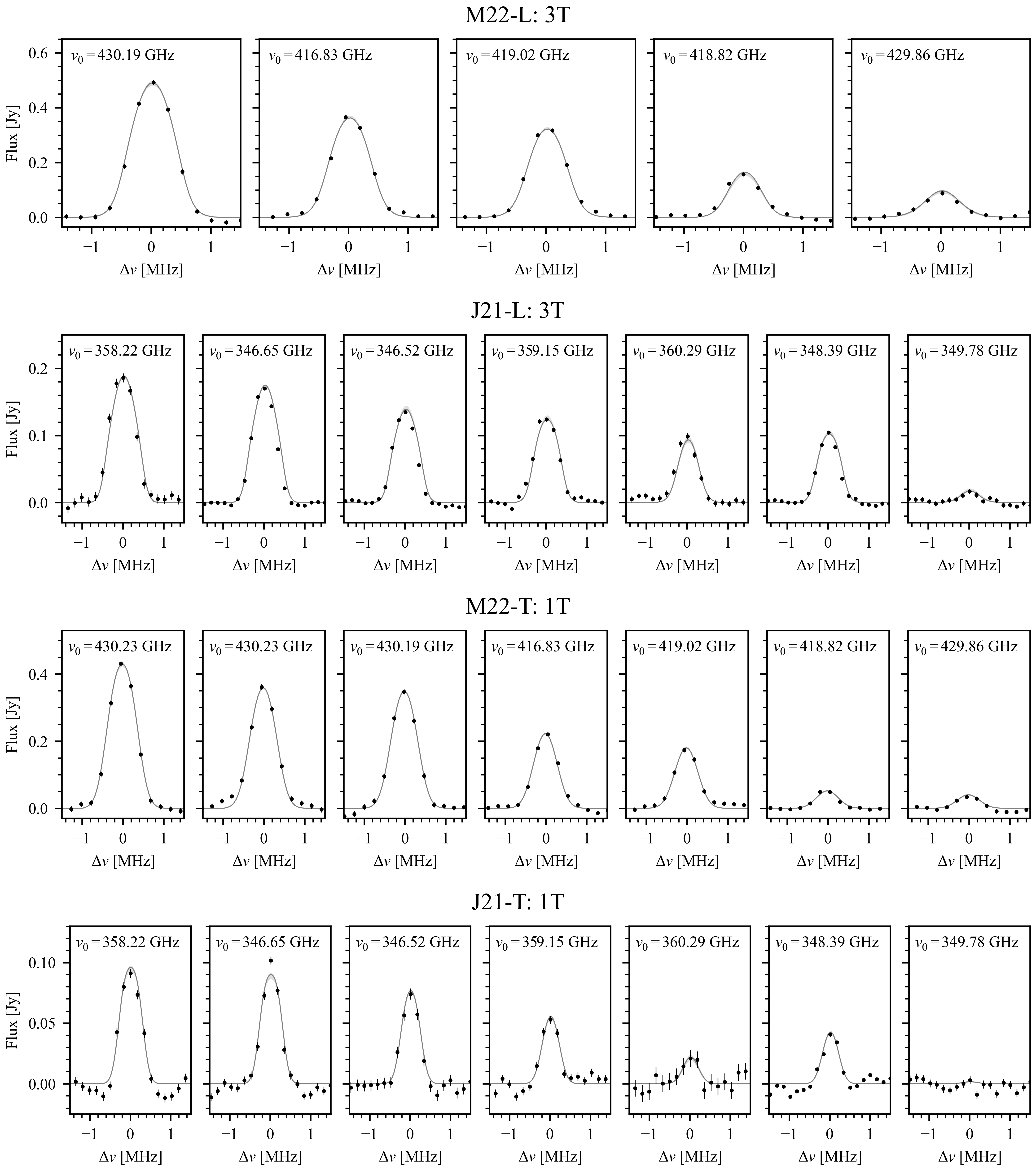}
    \caption{\textbf{Observed and modeled SO$_2$ rotational line spectra for all datasets.} Each row corresponds to a specific dataset and temperature profile parameterization, as labeled (M22-L: 3T, J21-L: 3T, M22-T: 1T, and J21-T: 1T). Individual panels show the targeted SO$_2$ transitions, which are labeled by their rest frequency $\nu_{0}$ and centered in frequency space. Observed spectra are plotted as black points. The best-fit model is overlaid as a dark gray line, with the lighter shading indicating the 1$\sigma$ uncertainty.}
    \label{fig:stacked_spectra}
\end{figure*}

\end{appendix}

\clearpage
\newpage
\bibliography{references}
\bibliographystyle{aasjournalv7}
\end{document}